\documentclass[a4paper,fleqn]{cas-sc}

\usepackage[authoryear]{natbib}
\usepackage{float}
\usepackage{lineno}
\def\tsc#1{\csdef{#1}{\textsc{\lowercase{#1}}\xspace}}
\tsc{WGM}
\tsc{QE}
\DeclareMathOperator{\arctantwo}{arctan2}
\begin{document}
\let\WriteBookmarks\relax
\def\floatpagepagefraction{1}
\def\textpagefraction{.001}

\shorttitle{Ab initio strewn field}    

\shortauthors{Carbognani, Fenucci, Salerno, Micheli}  

\title [mode = title]{Ab initio strewn field for small asteroids impacts}  


%\tnotetext[1]{<tnote text>} 


\author[1]{Albino Carbognani}[orcid=0000-0002-0737-7068]

%\cormark[1]


\ead{albino.carbognani@inaf.it}

\ead[url]{https://www.oas.inaf.it/}

\credit{Methodology, Computation}

\affiliation[1]{organization={INAF - Osservatorio di Astrofisica e Scienza dello Spazio},
            addressline={Via Gobetti 93/3 }, 
            city={Bologna},
            citysep={}, % Uncomment if no comma needed between city and postcode
            postcode={40129}, 
            country={Italy}}

\author[2, 3]{Marco Fenucci}[orcid=0000-0002-7058-0413]

% Footnote of the second author

% Email id of the second author
\ead{Marco.Fenucci@ext.esa.int}

\ead[url]{https://neo.ssa.esa.int/}

\credit{Study of NEOs orbits}

\affiliation[2]{organization={ESA ESRIN/PDO/NEO Coordination Centre},
            addressline={Largo Galileo Galilei, 1}, 
            city={Frascati (RM)},
            citysep={}, % Uncomment if no comma needed between city and postcode
            postcode={00044}, 
            country={Italy}}

\affiliation[3]{organization={Elecnor Deimos},
            addressline={Via Giuseppe Verdi, 6}, 
            city={San Pietro Mosezzo (NO)},
            citysep={}, % Uncomment if no comma needed between city and postcode
            postcode={28060}, 
            country={Italy}}

\author[4, 5]{Raffaele Salerno}[]

% Footnote of the third author
%\fnmark[4]

% Email id of the third author
\ead{raffaele.salerno@meteo.expert}

% URL of the third author
\ead[url]{https://www.meteo.expert/}

\credit{Computation of atmospheric profile}

\affiliation[4]{organization={Meteo Expert},
            addressline={Via G. Marconi, 27}, 
            city={Segrate (MI)},
            citysep={}, % Uncomment if no comma needed between city and postcode
            postcode={20054}, 
            country={Italy}}

\affiliation[5]{organization={Università Statale di Milano, Dipartimento di Scienze della Terra},
            addressline={Via S. Botticelli 23}, 
            city={Milano (MI)},
            citysep={}, % Uncomment if no comma needed between city and postcode
            postcode={20133}, 
            country={Italy}}

\author[2]{Marco Micheli}[orcid=0000-0001-7895-8209]

% Footnote of the fourth author

% Email id of the third author
\ead{Marco.Micheli@esa.int}

% URL of the third author
\ead[url]{https://neo.ssa.esa.int/}

\credit{Asteroids astrometry}

\cortext[1]{Corresponding author}

\fntext[1]{}



\begin{abstract}
In recent years, nine small near-Earth asteroids were discovered a few hours before the collision with the Earth: these are about one meter in diameter objects that have all disintegrated in the atmosphere, generating bright fireballs without causing damage. In some cases, several meteorites have been recovered. In cases like these, it is not always possible to triangulate the fireball generated by the asteroid's fall to circumscribe the strewn field position. For this reason, it can be important to compute a strewn field ``ab initio'', i.e. propagating the asteroid's trajectory in the atmosphere starting from the initial conditions obtained directly from the heliocentric orbit, coupled with some reasonable hypothesis about the mean strength and the mass of the fragments to ``sample'' the strewn field. By adopting a simple fragmentation model coupled with a real atmospheric profile, useful results can be obtained to locate the strewn field, as we will show for the recent falls of asteroids 2024~BX1, 2023~CX1 and 2008~TC3. It was possible to locate the strewn field of our study cases with an uncertainty of the order of one kilometre with the mean strength in the range 0.5-5 MPa and the mass of the possible final fragments in the 1 g - 1 kg range. We have also verified that a pancake phase after fragmentation is unnecessary to locate the strewn field for a small asteroid fall.

\end{abstract}

\begin{highlights}
\item Strewn field computation for small asteroid from orbital data only, without fireball triangulation.
\item Ab initio strewn field for 2023~CX1, 2024~BX1 and 2008 TC3 fall and comparison with recovered meteorites.
\item Fast determination of the strewn field of small asteroid for meteorite search. 
\item Possible strewn fields for 2022~WJ1 fall.
\end{highlights}

% Keywords
% Each keyword is separated by \sep

\maketitle

%%%%%%%%%%%%%%%%%%%%%%%%%%%%%%%%%%%%%%%%%%%%%%%%%%%%%%%%%%%%%%%%%%%%%%%%%%
\section{Introduction}
\label{sec:intro} 
In recent years, nine small near-Earth asteroids (NEAs), with a diameter of the order of one meter\footnote{According to the most recent definitions adopted by the International Astronomical Union, natural interplanetary objects with sizes between 30 micrometres and 1 meter are referred to as meteoroids \citep{IAU2024}. Despite this terminology, in the rest of the paper, we use the term asteroid for meter-sized objects because we are interested in bodies discovered with ground-based telescopes before impact and officially designated by the Minor Planet Center. Thus, they fell into the category of NEAs. In addition, the approach presented here is more of interest for application to NEAs; thus, we also use the term asteroid for future reference.}, have been discovered a few hours before hitting our planet, from 2008~TC3 on 7 October 2008, \citep[]{Farnocchia2017, shaddad-etal_2010}, to 2024~BX1 on 21 January 2024  \citep{Spurny2024}, ending with 2024~RW1 dropped in early September 2024, see Table~\ref{tab:small_impact} for a complete list.\\  
Asteroids of such small dimensions can be discovered thanks to the increased detection efficiency of NEA surveys over the last 20 years.  The discovery of small asteroids before they hit the Earth offers a good opportunity to understand the physical origin of these bodies compared to the more imprecise satellite observations \citep{Pena-Asensio2022}. Small NEAs of this size enter the atmosphere at hypervelocity and generally experience ablation, fragmentation and airburst, generating brilliant fireballs. The major fragments surviving the airburst can enter the dark flight phase and generate a fall event. In this case, several meteorites may be found on the so-called ``strewn field'' on the ground, allowing the chemical-physical characterisation of the parent body. Finally, knowledge of the orbit allows us to reconstruct, within certain limits, the dynamic history of the NEA. \\
Of the nine NEAs listed in Table~\ref{tab:small_impact}, in four cases (2008 TC3, 2018 LA, 2023 CX1 and 2024 BX1), meteorites were recovered; for 2014~AA, 2019~MO, 2022~EB5, 2022~WJ1 and 2024~RW1 the strewn field was placed in the water, and therefore no meteorites can be found. In reality, the case of 2022~WJ1 is borderline because the strewn field, based on the detection of falling fragments made by the Doppler radar KBUF located in Buffalo (NY), falls halfway between Lake Ontario and its south coast\footnote{\url{https://ares.jsc.nasa.gov/meteorite-falls/events/grimsby-ontario}}. We have treated this case separately in Appendix~\ref{sec:appendix}.\\
Of approximately 70\,000 meteorites recovered on Earth, only about 50 have their heliocentric orbits known, which can be determined using networks of all-sky cameras to triangulate fireballs, such as the Italian Prima Rete Italiana per la Sorveglianza sistematica di Meteore e Atmosfera\footnote{\url{http://www.prisma.inaf.it/}} \citep[PRISMA,][]{gardiol-etal_2016}, the French Fireball Recovery and InterPlanetary Observation Network\footnote{\url{https://www.fripon.org/}} \citep[FRIPON,][]{colas-etal_2020}, the European Fireball Network\footnote{\url{https://www.allsky7.net/}} \citep{oberst-etal_1998, Borovicka2022}, the Desert Fireball Network \citep[DFN,][]{Howie2017, Devillepoix2019} or the Spanish Meteor Network\footnote{\url{http://www.spmn.uji.es/}} \citep[SPMN,][]{trigo-rodriguez-etal_2001}, so every additional data is precious. The known meteorites originated from parent bodies of asteroidal origin, and there is no meteorite of certain cometary origin. Furthermore, a relevant part of meteorites with known orbit, about 25\%, may come from a parent body that originates directly in a collision between the NEAs population \citep{carbognani-fenucci_2023}.\\
In the case of small NEAs discovered while still in space and before hitting the Earth, the heliocentric orbit can be established from the astrometry obtained with ground-based telescopes, with higher precision than from fireball networks, and therefore, searching for a progenitor body is easier. However, if the fall is in an area not covered by an all-sky camera network, the main issue becomes locating the strewn field for meteorite search. The immediate recovery of meteorites is important because this minimises contamination from the terrestrial environment. Moreover, it is possible to detect the $\gamma-$rays emitted by radiogenic elements with half-lives down to a few days (${}^{47}\text{Ca}$, ${}^{52}\text{Mn}$, and ${}^{48}\text{V}$), generated by the exposure of the meteoroid to cosmic rays \citep{Bizzarri2023}. \\
Numerous fragmentation models can be used to describe the fall of a small asteroid into the atmosphere to locate the strewn field. One of the first models was developed by \cite{Passey1980} using classical drag and ablation equations on a flat Earth to simulate asteroid atmospheric breakup and study crater field formation. A later model is the ``pancake'', which is so-called because the impactor widens its radius continuously during the fall from a certain point onwards due to pressure from the shock wave. At a certain point, the fragments decouple, and each continues on its trajectory. This model was developed by \cite{Chyba1993} and \cite{Hills1993} and originally used to describe the Tunguska event or the airburst of a decametric asteroid in the atmosphere. In the models of \cite{Passey1980, Chyba1993, Hills1993}, a simple exponential law is used for the atmosphere's density considered to be isothermal. An alternative approach to describe asteroid fragmentation in the atmosphere is represented by ``discrete'' models, i.e. models that use a discrete fragment approach and treat the breakup as a series of fragmentation events that split the body into independent pieces as in \cite{Revelle2005}. In this model, the meteoroid divides into two identical child fragments, whose strength depends on the parent meteoroid's strength through a Weibull scaling law, producing a cloud of identical fragments at the end of the fragmentation process. A more recent model is the independent fragments and aggregate debris cloud, a fusion between a pancake and discrete model, called the Fragment Cloud Model (FCM), by \cite{Wheeler2017}. This hybrid model represents the breakup process as a series of progressive fragmentation events that split the meteoroid into several independent sub-fragments and a debris cloud in which the little fragments or dust are subject to common group aerodynamics. So, there is a pancake phase mixed with independent fragments simultaneously, and the fragments' speed is assumed to be equal to that of the parent body without a side component. The model can vary key parameters such as the number of fragments produced in each break, the mass distribution between fragments and pancake and fragment strengths to fit the observed energy deposition in the atmosphere during asteroid fall. In the FCM model, the atmospheric density is a curve fit of the 1976 standard atmosphere. \\
A model for the probabilistic assessment of asteroids' re-entry, named Asteroid Breakup Model (ABM), was developed by \cite{Limonta2021} using a modified version of the NASA Standard Breakup Model (SBM) used to generate the fragments distribution function in terms of their area-to-mass ratio and ejection velocity. The meteoroid is modelled as a point mass with uniform density and given area-to-mass ratio, subject to Earth gravity, air resistance and ablation. As is common with all these models, fragmentation occurs when the dynamic pressure equals the strength of the body. It is also assumed that no other breakup events will occur after the entire meteoroid undergoes fragmentation. The ABM describes the fragments cloud as a whole, i.e., it considers the fragments population as a fluid with continuous properties and propagates until they reach the ground or demise. This model was applied to the analysis of the 2008~TC3 fall, even if winds are not currently included, and the average distance between the theoretical and observed strewn field was about 2.2 km \citep{Limonta2021}.\\  
As for the location of the actual strewn fields, \cite{Ceplecha1987} was the first to describe the dark flight of meteoroids starting from the triangulation of ground cameras to compute strewn field position. More recently, a model used for computing the position of the strewn field starting from data given by the triangulation of all-sky networks is the dark flight Monte Carlo (DFMC) model that has successfully assisted in the meteorites recovery of Annama, the asteroid 2018 LA, and Ozerki \citep{Moilanen2021}. In this model, real wind profiles are included. Finally, a dark flight model based on the classical aerodynamic drag equation coupled with a real wind profile was used by the Desert Fireball Network (DFN) for meteorite recovery and is described by \cite{Towner2022}, using the Murrili fall as a case study. In this last case, the fireball's triangulation gives the starting dark flight point with speed and height, and the possible ground positions of meteoroids of different masses are calculated to sample the strewn field.\\
In this paper, we want to show that using a simple and fast-to-implement fragmentation model, a real winds profile and the NEA's orbital parameters as a starting point, we can determine the strewn field of small asteroids with good approximation. In this way, in case of a small asteroid discovered a few hours before impact, we can try to recover meteorites even in remote areas without dedicated networks for monitoring fireballs. To reach our goal, we will analyse the falls of two small asteroids discovered in space before hitting the Earth, 2023~CX1 and 2024~BX1, to which we added the historical case of 2008~TC3 as a further test.\\ 
We want to underline that we assume nothing about the meteoroid and the fireball apart from the initial atmospheric entry data derived from orbital elements and photometry. So, our aim is not to reproduce what was observed about the fireball but to show that using a simple fall model with reasonable starting parameters, the simulated strewn fields are in reasonable agreement with the observed ones. Therefore, it is possible to search for meteorites without necessarily having data from direct observations and triangulation of the fireball.\\
The paper is organised as follows: in Section~\ref{sec:model}, we will briefly outline our model used to describe the fall of small asteroids, discussing mass ablation, fragmentation, dark flight and strewn field; in Section~\ref{sec:2023CX1_2024BX1_event}, we will describe the 2023~CX1 and the 2024~BX1 events with the recovered meteorites and their mass and position on the ground. In Section~\ref{sec:astrometry_orbits}, we will see the astrometry and heliocentric orbit for our two near-Earth asteroids that give us the necessary starting condition for the fall model; in Section~\ref{sec:atmospheric_profile} we will present the atmospheric profile adopted in our case, while in Section~\ref{sec:2023CX1_strewn_field}, we will discuss the results about the strewn field given by the pancake model compared with the observed one, and the same in Section~\ref{sec:2024BX1_strewn_field} for 2024~BX1. In Section~\ref{sec:discussion}, there will be a discussion of our results, the critical points and the limitations, with a test about the 2008~TC3 strewn field. Finally, we provide our conclusions and Appendix~\ref{sec:appendix}, in which we report the results of our computations about the possible strewn fields of 2022~WJ1.

\begin{table*}
	\centering
	\caption{List of near-Earth asteroids discovered a few hours before the collision with Earth in order of date. The estimated diameter $D$ comes from the absolute magnitude value. The Minor Planet Electronic Circulars (MPEC) from the Minor Planet Centre are listed for each asteroid.}
	\label{tab:small_impact}
        \setlength\tabcolsep{2pt} % default value: 6pt
	\begin{tabular}{lcccllcc} 
		\hline
	Designation & Meteorites  & Impact date & UTC & Discoverer & Survey & Code & $D$ (m)\\
		\hline
2008 TC3$^{a}$  &  Y  & 2008-10-07   & 02:46 & Richard Kowalski    & Mount Lemmon Survey & G96 & 3-4     \\
2014 AA$^{b}$   &  N  & 2014-01-02   & 02:33 ($\pm 1$ h) & Richard Kowalski    & Mount Lemmon Survey & G96 & 3     \\
2018 LA$^{c}$   &  Y  & 2018-06-02   & 16:44 & Richard Kowalski    & Mount Lemmon Survey & G96 & 3-4   \\
2019 MO$^{d}$   &  N  & 2018-06-22   & 21:25 &       ---           & ATLAS-MLO           & T08 & 4-6   \\
2022 EB5$^{e}$  &  N  & 2022-03-11   & 21:22 & Krisztián Sárneczky & Konkoly Observatory & K88 & 2     \\
2022 WJ1$^{f}$  &  N  & 2022-11-19   & 08:27 & David Rankin        & Mount Lemmon Survey & G96 & 1     \\
2023 CX1$^{g}$  &  Y  & 2023-02-13   & 02:59 & Krisztián Sárneczky & Konkoly Observatory & K88 & 1     \\
2024 BX1$^{h}$  &  Y  & 2024-01-21   & 00:32 & Krisztián Sárneczky & Konkoly Observatory & K88 & 1     \\
2024 RW1$^{i}$  &  N  & 2024-09-04   & 16:39 & Jacqueline Fazekas  & Mount Lemmon Survey & G96 & 1.5     \\
		\hline
  \\
  \multicolumn{8}{l}{$^{a}$ MPEC 2008-T72, \url{https://minorplanetcenter.net/mpec/K08/K08T72.html}}\\
  \multicolumn{8}{l}{$^{b}$ MPEC 2014-A02, \url{https://minorplanetcenter.net/mpec/K14/K14A02.html}}\\
  \multicolumn{8}{l}{$^{c}$ MPEC 2018-L04, \url{https://minorplanetcenter.net/mpec/K18/K18L04.html}}\\
  \multicolumn{8}{l}{$^{d}$ MPEC 2019-M72, \url{https://minorplanetcenter.net/mpec/K19/K19M72.html}}\\
  \multicolumn{8}{l}{$^{e}$ MPEC 2022-E178, \url{https://minorplanetcenter.net/mpec/K22/K22EH8.html}}\\
  \multicolumn{8}{l}{$^{f}$ MPEC 2022-W69, \url{https://minorplanetcenter.net/mpec/K22/K22W69.html}}\\
  \multicolumn{8}{l}{$^{g}$ MPEC 2023-C103, \url{https://minorplanetcenter.net/mpec/K23/K23CA3.html}}\\
  \multicolumn{8}{l}{$^{h}$ MPEC 2024-B76, \url{https://minorplanetcenter.net/mpec/K24/K24B76.html}}\\
  \multicolumn{8}{l}{$^{i}$ MPEC 2024-R68, \url{https://minorplanetcenter.net/mpec/K24/K24R68.html}}\\
	\end{tabular}
\end{table*}

\section{The fall model}
\label{sec:model}
As a fall model, we started from that described in \cite{carbognani2024}, regarding a possible strewn field of the Tunguska event using a pancake model, removing the pancake phase, reasonable for decametric asteroids but negligible for a meter body, and considering only the fragmentation. Although this model might seem rather obsolete compared to more sophisticated variants such as \cite{Wheeler2017}, we will see that it can provide reasonable predictions about the possible position of the strewn field. In the following, we briefly review the main characteristic of our model, which differs from the originals of \cite{Chyba1993, Hills1993} because the asteroid’s motion is referenced to the geocenter, and the strength of the fragments is considered higher than that of the original body.\\ 
When a small asteroid falls into the terrestrial atmosphere, the path can be qualitatively divided into three main phases. The first is when the asteroid enters the upper atmosphere with a speed of the order of 10-20 km/s; the drag starts and a shock wave is formed in the front of the body, the air is compressed and heated, and mass ablation begins: this is the fireball phase. The second phase starts when the pressure of the shock wave is higher than the body's strength, and fragmentation occurs. In case of fragmentation into many small fragments and dust particles, the body's mass spreads over a greater area, the quantity of intercepted atmosphere increases, and thus, braking and ablation increase: the body loses kinetic energy very rapidly, i.e. explosively, and there is a little airburst. In the case of fragmentation into large fragments, the airburst is less conspicuous, and there is no airburst in rare cases without fragmentation. Afterwards, the fragments decouple, developing their shock wave as independent fireballs. \\     
Finally, the dark flight phase of each fragment falling into the atmosphere begins when the atmospheric speed drops below about 03 km/s \citep{Passey1980, Ceplecha1987}. This is the end of the fireball phase. In the dark flight phase, the original cosmic speed of the body is completely lost to drag, and fragments enter the densest part of the terrestrial atmosphere; therefore, the wind speed and direction become very important to determine the impact point with the terrestrial surface, i.e. the strewn field to search for meteorites: in the dark flight phase, an atmospheric wind profile from about 30 km to the ground is necessary.\\
From elementary physics, the classical equation of motion in a geocentric inertial reference system describing the fall of a meteoroid with mass $m$ and acceleration $\vec{a}_m$ in the dark flight phase is as follows:

\begin{equation}
m\vec{a}_m =  \vec{F}_g + \vec{F}_d = -GMm\frac{\vec{r}}{r^3} -\Gamma\rho_a \left|\vec{V}_m - \vec{W}\right| A\left( \vec{V}_m - \vec{W} \right)
\label{eq:motion_inertial}
\end{equation}

\noindent Eq.~(\ref{eq:motion_inertial}) is a non-linear differential equation in vector form. The first term on the right is the gravity force ($G$ is the gravitational constant, $M$ is the Earth's mass and $\vec{r}$ the distance between the geocenter and the meteoroid), the second is the drag force (Newton's Resistance law) exerted by the air on the meteoroid: $\Gamma$ is the dimensionless drag coefficient (equal to 0.5 for a sphere), $\rho_a$ the air density, $A$ the meteoroid cross section, $\vec{V}_m$ the meteoroid speed and $\vec{W}$ the wind speed. The value of the drag coefficient $\Gamma$ depends both on the unknown final form of the meteoroid after ablation and on the Mach number, i.e., the ratio between the meteoroid speed and the sound speed at the same height above ground \citep{Ceplecha1987}. The value of the drag coefficient is independent of the size, the crucial parameter being the body shape. The $\Gamma$ asymptotic value, i.e., toward very high Mach numbers, is left as a free parameter. In contrast, for low Mach numbers, i.e. equal or less than 4, we adopt the following Ceplecha's values: $\Gamma(4)=0.58$, $\Gamma(3)=0.62$, $\Gamma(2)=0.63$, $\Gamma(1)=0.50$, $\Gamma(0.8)=0.44$, $\Gamma(0.6)=0.39$, $\Gamma(0.4)=0.35$ and $\Gamma(0.2)=0.33$. Instead of using a reference system with the axes fixed in space, we used a geocentric reference system with the axes rotating with the Earth's surface, where the observer is located. In this case, two inertial forces must be added: the Coriolis and centrifugal. These two inertial forces are generally negligible about the strewn field location, but we include them for completeness. \\
In the fireball phase, in addition to the vector equation describing atmospheric braking, it is also necessary to consider the scalar mass loss equation describing the thermal ablation process \citep{Chyba1993}: 

\begin{equation}
\frac{d m}{dt}= -\Gamma \frac{C_H}{Q} \rho_a A {{V}_m}^3
\label{eq:mass_loss_ablation}
\end{equation}

\noindent In Eq.~(\ref{eq:mass_loss_ablation}) $Q$ is the heat of ablation (a combination of the heat of fusion and the heat of vaporisation \citep{Passey1980}), $C_H$ is the heat transfer coefficient, and the drag coefficient $\Gamma$, in this case, is a constant because this equation intervenes at hypersonic speeds only. For stony asteroids $Q\approx 8\cdot 10^6$ J/kg, while $C_H\approx 0.1$, although the value of the heat transfer coefficient tends to decrease for altitudes lower than 30 km \citep{Chyba1993, Avramenko2014, Johnston2018}. Note that the adopted standard values for $Q$ and $C_H$ are not necessarily the right choices. For example, arc jet experiments suggest that the ablation heat was about 4 MJ/kg \citep{Stern2017}, while for the heat transfer coefficient, a lower value of around 0.05 might be more correct as suggested by \citet{Johnston2018}.
If the meteoroid's mass decreases, the value of section $A$ also decreases accordingly. As we said at the beginning, ablation ceases when the velocity drops below approximately 03 km/s. Solving these equations numerically with a Runge-Kutta 4th/5th order solver and using the World Geodetic System 84 (WGS 84) provides the position and velocity of the meteoroid as it falls towards the ground, and the intersection of the trajectory with the ground gives the impact point. \\ 

\noindent Usually, the meteoroids fragmentation model assumes that the process starts when the aerodynamic pressure from Eq.~(\ref{eq:motion_inertial}) $P_{\textrm{dyn}}=\Gamma \rho_{\textrm{fr}} {V_m}^2$ in front of the body is equal or superior to mechanical strength $S$ of the body: $P_{\textrm{dyn}}\geq S$ \citep{Chyba1993}. The quantity $\rho_{\textrm{fr}}$ is the air density at the fragmentation height, $V_m$ is the body speed relative to the air, and $\Gamma$ is the dimensionless drag coefficient. \\
Let's suppose the mechanical destruction of a monolithic body, as is reasonable to expect for small asteroids, occurs along the fracture lines. In that case, it is reasonable to expect that the strength depends on the volume or mass and that the following scale relation holds \citep{Weibull1951, Svetsov1995, Scheeres2015}:

\begin{equation}
S_{main} = S_{fr}\left( \frac{m_{fr}}{m_{main}} \right)^{\alpha}
\label{eq:scale_strength}
\end{equation}

\noindent This law of scaling, based on Weibull's statistic, derives from the fact that if the mass decreases, the size of the weak points in the rock must also decrease, so the strength must increase. In Eq.~(\ref{eq:scale_strength}) $\alpha$ is the Weibull modulus, $S_{main}$ is the main body strength while $S_{fr}$ is the fragment strength. The $\alpha$ value increases as the inhomogeneity of the material increases, and a value between 0.1 and 0.7 is expected. We assume $\alpha\approx 0.2$ in our model, as for Chelyabinsk \citep{carbognani2024}. So, when the small asteroid disrupts, the resulting fragment, statistically, has a strength superior to that of the original body and can survive for the next path phases. This happens in our case: after the first fragmentation, the fragments always have a strength greater than the pressure exerted by the atmosphere, and there are no further fragmentations. This is a simplification, considering that multiple fragmentations were observed in some fireballs. However, as we will see, even with a single fragmentation, the results obtained for the strewn fields of known cases are good and would allow a meteorite search.\\
A priori, we do not know the masses of the fragments in the strewn field, i.e. the meteorites' masses. So, in our model, we assume the following final masses with spherical shape: 1, 0.3, 0.2, 0.1, 0.05, 0.02, 0.005 and 0.001 kg. In practice, we use some ``sample'' particles to show the final strewn field's possible range and position. With good approximation, these masses for the fragments correspond to the typical masses of meteorites on the ground. In all cases, the sum of the fragment's mass is inferior to the meteoroid mass after the fragmentation phase, so we assume that the great part of the asteroid's mass was lost in ablation and debris cloud. Note that after the airburst, there is a residual ablation phase, which tends to decrease the mass of the fragments, which we consider in the computation. So, the original fragment mass after the fragmentation is larger than our assumed meteorite's mass that ``samples'' the strewn field. There is no guarantee that any meteorites corresponding to these assumed masses will be produced, even if they are the mass of typical meteorites. \\
When a meter-sized asteroid enters the atmosphere, various outcomes can occur. In rare cases, the body will be monolithic, and no fragmentation will occur; an example was the Carancas event in 2007 \citep{Borovicka2008}. A complete disintegration with no meteorite, as could happen for bodies of cometary origin, is more unlikely. Still, this eventuality must be kept in mind based on the orbit type of the impacting body. Anyway, for these fragments, we will assume that the speed is equal to that of the fragmentation phase, with the same inclination and direction as the original body. With this last assumption, we simplified the model because we neglected the possible lateral expansion speed imprinted during fragmentation. This speed component is usually very small, so pieces from a fragmented fireball, in most cases, continue along their original trajectory \citep{Moilanen2021}. In our picture, all the fragments were subject to acceleration along the trajectory given by Eq.~(\ref{eq:motion_inertial}), and considering that the mass-area ratio $m/A$ is proportional to the fragment's radius, the smaller the meteoroid and greater the negative drag acceleration will be. This caused smaller fragments to fall earlier than the bigger fragments.

\subsection{Is there a pancake phase for a small asteroid fall?}
\label{sec:pancake_phase}
The applicability of the pancake model to the fall of small meteoroids is an open question because the model was originally proposed to explain the airburst of decametric bodies (such as Tunguska) producing large amounts of fragments, which can have a common shock wave, factors that would be missing in a smaller body. However, it cannot be ruled out that even small asteroids of the order of a meter in diameter could generate a short pancake phase due to the possible fragility of the surface exposed to extreme space conditions such as temperature variations, bombardment by micrometeoroids and cosmic rays. For example, in \cite{Wheeler2018}, the FCM model with pancake and separate fragments, briefly described in Section~\ref{sec:intro}, is successfully used to model the energy loss of a group of small meteoroids with meteorites fall such as Košice (1.4 m diameter), Benešov (1.3 m in diameter) and Lake Tagish (4.5 m in diameter). \\
For this reason, we compared the results obtained from the model with fragmentation alone and the one with fragmentation followed by a short pancake phase. The equations used to represent the pancake phase are the following. From general dimensional considerations, equating the expansion work made during the pancake phase with the kinetic energy of expansion of the debris, the post-breakup dispersal speed, $V_{disp}$, of the meteoroid fragments in the pancake phase is given by \citep{Hills1993}:

\begin{equation}
V_{disp} \approx \sqrt{\frac{7}{2}\frac{\rho_a}{\rho_m}}V_m
\label{eq:dispersal_speed}
\end{equation}

\noindent This is a general equation also used in \citet{Wheeler2017}. In Eq.~(\ref{eq:dispersal_speed}) $\rho_m$ is the meteoroid mean density. If we fixed the value of $V_m$, the lateral speed increases as the air density increases; therefore, if the fragmentation occurs at a low altitude, the lateral expansion speed is greater. For a typical fragment's mean density of $\rho_m\approx 3000~\text{kg m}^{-3}$, a typical atmospheric speed of $V_m\approx 15$ km/s and an atmospheric density $\rho_a\approx 0.02~\text{kg m}^{-3}$ (corresponding to a height of about 30 km), the dispersal speed is about the 0.5\% of the original speed, with a deflection of about $0.3^\circ$ respect to the original trajectory. The diameter of the pancake structure from the break time $t=0$ is:

\begin{equation}
D \approx D_0 + 2\int_{0}^{t} V_{disp}dt
\label{eq:pancake_diameter}
\end{equation}

\noindent This diameter growth over time increases the drag of the fragmented asteroid and causes a sudden decrease in speed. When the diameter of the pancake structure becomes larger than a certain number of times the original diameter, it is assumed that the fragments decouple and that each fragment continues to fall as an independent body with its shock wave.  \\
In experiments with groups of spheres of identical diameter coupled together, it was found that the decoupling is complete when the diameter of the group of spheres that move away is about 2.28 times the original one \citep{Whalen2021} therefore, the ratio between the initial diameter and the final diameter of the pancake phase must be at least greater than this value. However, the further fragmentation that occurs during the pancake phase can delay the decoupling between the fragments so that a 5-10 times aspect ratio between the final diameter of the pancake phase and the original diameter (the so-called ``pancake factor'', which we will indicate with the letter $k$) can correctly describe the airburst altitude \citep{Chyba1993, Collins2005}. It should be noted that high pancake factor values, although empirically providing correct values for the airburst, are not considered physically realistic because - if interpreted literally - the pancake structure would become very thin compared to the diameter \citep{Collins2017}. As a reference value for the pancake factor, we will assume $k\approx 3.5$. \\

\subsection{A summary}
\label{sec:summary_list}
To summarise, the adopted model for the description of the fall of small NEAs (with the designation of the Minor Planet Center) has eight parameters: (1) the heat of ablation $Q$; (2) the heat transfer coefficient $C_H$; (3) the drag coefficient $\Gamma$, whose values we have already discussed which determine atmospheric ablation and braking; (4) the mean strength $S$ of the asteroid, which is very important because determines the height of the fragmentation processes, the length of the path followed in the atmosphere and, together with the wind direction and intensity, the strewn field position; (5) the Weibull modulus $\alpha$, which determines the fragment's strength and (6) the pancake factor (a parameter absent in the model with fragmentation only) which determines the fragments decoupling at the end of the pancake phase. The greater the duration of this phase, the greater the supplementary drag due to increased pancake dimension; (7) the drag and the ablation that the asteroid experiences in the atmosphere depend on the cross-section $A\approx\pi D^2$, where the diameter $D$ is an important parameter to know a priori. Assuming the albedo of the asteroid, it can be estimated from the absolute magnitude $H$ value determined by telescopic observations before the fall; (8) the last parameter is the mean asteroid density $\rho_m$ that, with the diameter $D$, determines the mass value $m$. The parameters $Q\approx 8\cdot 10^6$ J/kg, $Q_H\approx 0.1$, $\alpha\approx 0.2$, and $k\approx 3.5$ will be kept fixed at the values indicated in the text, while the strength $S$ will be a parameter that can vary. \\
In the case of the major fireballs observed by the European Fireball Network, the strength of the first fragmentation is in the 0.4-4 MPa range while the main fragmentation is in the 3.5-12 MPa range \citep{Borovicka2008}. In a later work involving ordinary chondritic meteoroids \citep{Borovicka2020}, the first fragmentation typically corresponds to a low strength of 0.04–0.12 MPa. In two-thirds of cases, the first phase was catastrophic or near-catastrophic, with the loss of at least 40\% of the mass. The second fragmentation corresponds to 0.9–5 MPa. \\
In our model, as it is constructed, we only have one fragmentation, and we use some test values for the strength: the first with $S\approx 1$ MPa and the second with $S\approx 5$ MPa, with an extension toward $S\approx 0.5$ MPa in some cases. This will allow us to see the effect of the average strength on the position of the theoretical strewn field compared to the real one. The atmospheric profile will be computed around the starting position of the theoretical dark flight with $S\approx 1$ MPa. Finally, the heliocentric orbit of the body will give the starting latitude and longitude, the direction of the motion, trajectory inclination to the local surface and speed at the top of the atmosphere (which we assume is 100 km from the ground), considering the effect of the Earth's gravity.

\section{The 2023 CX1 and 2024 BX1 impact events}
\label{sec:2023CX1_2024BX1_event}
On the evening of 12 February 2023, at 20:18 UTC, Hungarian astronomer Krisztián Sárneczky discovered an object of magnitude +19.4 with the 60 cm Schmidt telescope of the Piszkéstető Mountain Station (IAU K88), an observatory of the Hungarian Academy of Sciences located about 80 km north-east of Budapest. The object was immediately included in the NEO Confirmation Page (NEOCP) of the Minor Planet Center (MPC) for confirmation and follow-up observations by other observatories with the acronym Sar2667.\\
During the follow-up, the ESA Meerkat and JPL Scout systems indicated that the object was on a collision course with Earth and would fall into the English Channel near the coast of Normandy at about 03 UTC on 13 February. More than 250 astrometric measurements were collected during the 6.5 hours between discovery and impact, and at 04:13 UTC on 13 February MPEC 2023-C103\footnote{\url{https://minorplanetcenter.net/mpec/K23/K23CA3.html}} was released by the MPC, assigning the designation 2023 CX1 to the newly fallen asteroid. \\
The International Meteor Organization (IMO) received 84 eyewitness reports about the fireball associated with 2023 CX1 (Event 937-2023), which was also recorded on videos and pictures by numerous people waiting for the last asteroid appearance\footnote{\url{https://www.imo.net/2023-cx1-7th-predicted-earth-impact/}}. 
Unfortunately, there were no multiple detections by the FRIPON all-sky camera network. An approximate strewn field calculated by Peter Jenninskens (SETI – USA), Denis Vida (UWO, Canada), Auriane Egal (UWO and Space for Life, Montreal) and Hadrien Devillepoix (DFN – Australia) showed that meteorites would have impacted in an area between Dieppe and Doudeville in Normandy. The first city is about 20 km east of the true strewn field; the second city is about 11 km south-west, as we see later\footnote{\url{https://www.fripon.org/meteorites-found-after-observation-of-asteroid-2023cx1-fireball-above-normandy-france/}}. Also Jiří Borovička and Pavel Spurný (Astronomical Institute of the Czech Academy of Sciences) used the publicly available images and videos of the fireball to triangulate the trajectory and compute a strewn field\footnote{\url{https://www.imo.net/the-atmospheric-trajectory-of-2023-cx1-and-the-possible-meteorite-strewn-field/}}. Based on light curves, a destructive fragmentation at about 28 km height occurs, and the strewn field, computed with the approximate wind profile from Herstmonceux (UK), was located between Angiens and Houdetot, along the true strewn field. Unfortunately, a paper on fireball triangulation associated with 2023 CX1 hasn't been published yet.\\
Based on what is reported in the Meteoritical Bulletin Database\footnote{\url{https://www.lpi.usra.edu/meteor/metbull.php?code=79565}}, the FRIPON/Vigie-Ciel project teams arrived on-site on the day following the fall. The first systematic search was conducted on 15 February, and in the afternoon of the same day, the first stone (94 g) was found in the commune of Saint-Pierre-le-Viger by art student Loïs Leblanc-Rappe. The search continued during the following days, leading to the recovery of eleven more small stones (mass range 2-24 g) in the commune of Angiens. Fieldwork for the FRIPON/Vigie-Ciel teams ended on 26 March 2023. The strewn field has an estimated size of about $8\times 3$ km, but the real strewn field may be larger. The meteorite Saint-Pierre-le-Viger is classified as an Ordinary Chondrite L5-6 \citep{Bischoff2023}; see Table~\ref{tab:meteorites_2023CX1} for a list of the recovered meteorites with the associate position.\\

\begin{table*}
	\centering
	\caption{Official meteorites list from 2023 CX1 fall, with mass value and recovery latitude and longitude. Data from the Meteoritical Bulletin Database.}
	\label{tab:meteorites_2023CX1}
        \setlength\tabcolsep{2pt} % default value: 6pt
	\begin{tabular}{llll} 
\hline
Mass name &	    	Mass (g)	& Lat. N 	& Long. E  \\
\hline
Saint-Pierre-le-Viger [1]	&	94    &	49°49'15" &	0°49'35" \\	
Saint-Pierre-le-Viger [2]	&	3     & 49°49'58" &	0°46'54" \\		
Saint-Pierre-le-Viger [3]	&	7.6   &	49°49'53" &	0°47'10" \\ 	
Saint-Pierre-le-Viger [4]	&	9     &	49°49'59" &	0°47'10" \\  		
Saint-Pierre-le-Viger [5]	&	7.2   & 49°49'58" &	0°47'0"	 \\ 
Saint-Pierre-le-Viger [6]	&	4     & 49°49'56" &	0°47'4"	 \\ 	
Saint-Pierre-le-Viger [7]	&	5.4   &	49°49'52" &	0°47'11" \\	
Saint-Pierre-le-Viger [8]	&	11    &	49°49'33" &	0°47'43" \\	
Saint-Pierre-le-Viger [9]	&	17    &	49°49'23" &	0°47'54" \\	
Saint-Pierre-le-Viger [10]	&	23.6  &	49°49'32" &	0°47'51" \\	
Saint-Pierre-le-Viger [11]	&	23.8  &	49°49'22" &	0°48'0"	 \\
Saint-Pierre-le-Viger [12]	&	2     & 49°49'27" &	0°48'2"	 \\ 	
Saint-Pierre-le-Viger [13]	&	175.2 &	49°48'29" &	0°49'41" \\ 	
Saint-Pierre-le-Viger [14]	&	490   & 49°47'46" &	0°51'35" \\ 

		\hline
	\end{tabular}
\end{table*}

\noindent About 2024 BX1, the first detection was also obtained by Sárneczky, who discovered the asteroid at 21:48 UTC on 20 January 2024, with an apparent magnitude of +18.0. The new object was included in the NEOCP with the acronym Sar2736, and confirmatory observations from European observers immediately began. The early impact warning software Scout at JPL, Meerkat at ESA and the NeoScan system by NEODyS immediately warned of a high impact probability with the Earth. The asteroid fell at 00:32 UTC on 21 January 2024, entering the atmosphere about 60 km west of Berlin, near Nennhausen. About an hour after the event, the MPC published MPEC 2024-B76\footnote{\url{https://www.minorplanetcenter.net/mpec/K24/K24B76.html}} that assigned the designation 2024 BX1 to the asteroid that just fell, and published almost 200 astrometric observations of the object. The absolute magnitude is $H\approx 32.8$ equivalent, for a geometric albedo in the range 0.05-0.25, to an object with a diameter of 1-2 m.\\
Thanks to the fireball detection from the European Fireball Network and also with data from the AllSky7 network, it was possible to triangulate the trajectory of the fireball, which lasted approximately 5 seconds with a peak absolute magnitude of -14.4 at 34$-$35 km heights. Significant fragmentations occurred between 38–30 km heights, but many other secondary events occurred between 54-29 km \citep{Spurny2024}. The terminal point of the fireball and the start of the dark flight phase was located at the height of 21.2 km at coordinates Lat. $52.635^\circ$ N, Long. $12.642^\circ$ E. From the fireball model, an entry mass of about 140 kg was estimated, equivalent to a diameter of about 0.44 m: a value lower than expected based on absolute mag value. This discrepancy was probably due to the higher albedo of the asteroid than expected. According to \cite{Spurny2024}, the strewn field was located 10 kilometres west of the German town of Nauen, in the Land of Brandenburg. The ground is flat in the area, rich in cultivated fields, and it was not difficult to recover meteorites: on 25 January, a commercial Polish search team found the first three meteorites; the next day, researchers from the Berlin Museum of Natural Sciences, the Free University of Berlin and the German Aerospace Center recovered other meteorites. There were many participants in the search, and the position was not always reported, but overall, 202 meteorites were collected for a total mass of 1.8 kg, while the observed strewn field has an extension of 1.5 $\times$ 10 km \citep{Bischoff2024}. According to the Meteoritical Bulletin Database, the meteorites associated with this event are classified as an Aubrite, and the official name is Ribbeck\footnote{\url{https://www.lpi.usra.edu/meteor/metbull.php?code=81447}}. In Table~\ref{tab:meteorites_2024BX1}, we have reported a partial meteorites list; a complete list is available in the supplementary material of \cite{Bischoff2024}.\\

\begin{table*}
	\centering
	\caption{Partial meteorites list from 2024 BX1 fall, with mass value, latitude and longitude in recovery order. Data from the Karmaka Meteorites website (\url{https://karmaka.de/?p=34832}), last update 29 May 2024. A more complete list is given in \cite{Bischoff2024}, which we used for the strewn field plot in Fig.~\ref{fig:2024BX1_strewn_field}. Ribbeck is a hamlet in the German city of Nauen.}
	\label{tab:meteorites_2024BX1}
        \setlength\tabcolsep{2pt} % default value: 6pt
	\begin{tabular}{llll} 
\hline
Mass name &	Mass (g) & Lat. N 	& Long. E  \\
\hline
Ribbeck	[1] &	171.66   & 	52°37'38.1" &	12°43'50.6" \\		
Ribbeck	[2] &	111.193  &	52°37'23.7" &	12°44'21.6" \\ 	
Ribbeck	[3] &	52.36    &	52°37'32.5" &	12°44'42.1" \\  		
Ribbeck [4] &	45.89    & 	52°37'07.8" &	12°45'56.9" \\ 
Ribbeck	[5] &	20.51    & 	52°37'05.8" &	12°45'53.6" \\ 	
Ribbeck	[6] &	15.59    &	52°36'54.2" &	12°46'21.4" \\	
Ribbeck	[7] &	14.20    &	52°37'05.3" &	12°45'33.3" \\	
Ribbeck	[8] &	10.20    &	52°36'55.5" &	12°46'24.8" \\	
Ribbeck [9] &	4.77     &	52°36'35.9" &	12°47'48.7" \\	
Ribbeck [10] &	26.2     &	52°37'03.5" &	12°46'11.8" \\
Ribbeck [11] &	9.251    &  52°37'15.1" &	12°45'40.1" \\ 	
Ribbeck [12] &	3.384    &	52°36'41.8" &	12°46'39.8" \\ 	
Ribbeck [13] &	5.20     &  52°36'55.4" &	12°46'29.5" \\ 
Ribbeck [14] &	225.0    &	52°37'26.7" &	12°44'23.8" \\ 	
Ribbeck [15] &	10.62    &  52°36'34.308" &	12°47'21.681" \\ 
Ribbeck	[16] &	21.0     & 	52°36'30.380" &	12°47'21.257" \\		
Ribbeck	[17] &	9.2      &	52°36'28.1"   &	12°47'46.0" \\ 	
Ribbeck	[18] &	8.0      &	52°36'23.0"   &	12°48'23.4" \\  		
Ribbeck [19] &	11.11    & 	52°36'48.6"   &	12°47'00.6" \\ 
Ribbeck	[20] &	5.45     & 	52°36'19.3"   &	12°48'28.2" \\ 	
Ribbeck	[21] &	5.70     &	52°36'24.8"   &	12°48'16.9" \\	
Ribbeck	[22] &	2.86     &	52°36'46.2"   &	12°48'27.1" \\	
Ribbeck	[23] &	11.7     &	52°36'42.0"   &	12°47'04.0" \\	
Ribbeck [24] &	4.14     &	52°36'37.68"  &	12°48'30.78" \\	
Ribbeck [25] &	5.67     &	52°36'14.4"   &	12°48'42.8" \\
Ribbeck [26] &	3.73     &  52°36'43.6"   &	12°48'14.3" \\ 	
Ribbeck [27] &	8.11     &	52°36'54.0"   &	12°47'33.0" \\ 	
Ribbeck [28] &	2.03     &  52°36'30.168" &	12°49'08.327" \\ 
Ribbeck [29] &	47.5     &	52°36'32.7"   &	12°47'42.8" \\ 	
Ribbeck [30] &	7.79     &  52°36'29.2"   &	12°48'17.1" \\ 
Ribbeck	[31] &	2.4      & 	52°36'23.0"   &	12°49'19.9" \\		
Ribbeck	[32] &	2.38     &	52°36'27.3"   &	12°49'24.4" \\ 	
Ribbeck	[33] &	2.86     &	52°36'27.1"   &	12°48'10.6" \\  		
Ribbeck [34] &	3.01     & 	52°36'15.552" &	12°49'47.1" \\ 
Ribbeck	[35] &	4.48     & 	52°36'32.256" &	12°47'03.66" \\ 	
Ribbeck	[36] &	7.36     &	52°36'32.3"   &	12°48'31.2" \\	
Ribbeck	[37] &	1.56     &	52°36'32.7"   &	12°47'42.8" \\	
Ribbeck	[38] &	1.49     &	52°36'22.9"   &	12°49'22.9" \\	
Ribbeck [39] &	0.872    &	52°36'32.0"   &	12°49'02.1" \\	
Ribbeck [40] &	3.335    &	52°36'19.44"  &	12°50'18.66" \\

		\hline
	\end{tabular}
\end{table*}

\section{Astrometry and orbits of 2023 CX1 and 2024 BX1}
\label{sec:astrometry_orbits}

% Orbit determination
The orbit determination used in this work has been performed with the ESA Aegis software \citep{faggioli-etal_2023}. 
The orbital elements are determined through a least-square method aimed at minimizing the sum of the squared residuals. The residuals are defined as the difference between the measured astrometric positions and the ones computed from a dynamical model. The lest-square problem is solved by using differential corrections \citep{milani-gronchi_2009}, that is endowed with an automated outlier rejection scheme \citep{carpino-etal_2003} aimed at identifying observations not compatible with the orbit within the assumed astrometric errors. 
The dynamical model used for orbit fitting includes the gravitational perturbations of the Sun and the eight planets from Mercury to Neptune, the 16 most massive main-belt asteroids, and Pluto \citep[see also][for values of their masses]{fenucci-etal_2024}. Their positions and velocities are retrieved from the JPL Planetary and Lunar Ephemeris DE441 \citep{park-etal_2021}. The oblateness terms $J_2$ of the Earth and the Sun are also taken into account, and their relativistic contributions are included by using the post-Newtonian formalism \citep{will_1993}. 

% Orbital elements table
\begin{table}[!ht]
    \centering
    \caption{Keplerian orbital elements of 2023~CX1 and 2024~BX1, with the corresponding epoch expressed in MJD. Errors refer to the 1-$\sigma$ formal uncertainties.}
    \begin{tabular}{ccc}
         \hline
      Object                 & 2023~CX1                                   & 2024~BX1  \\
         \hline
      Epoch (MJD)            & 59988.07747 TDT                        &  60329.98788 TDT          \\ 
      Semi-major axis (au)        &  $1.75811384         \pm 0.000016$      &  $1.38815446   \pm  0.000020$  \\ 
      Eccentricity           &  $0.47987242         \pm 0.0000054$        &  $0.40559618   \pm  0.000011$     \\ 
      Inclination (deg)           &  $3.78607632   \pm 0.000041$   &  $7.83713738   \pm  0.00021$      \\ 
      Longitude of node (deg)      &  $323.82613574 \pm 0.00000068$ &  $300.10735473 \pm  0.00000071$   \\ 
      Argument of perihelion (deg) &  $219.40492969 \pm 0.000046$   &  $243.88337381 \pm  0.00011$      \\ 
      Mean anomaly  (deg)         &  $347.17285000 \pm 0.00020$    &  $332.02689986 \pm  0.00080$      \\ 
      Normalized RMS         & 0.400  & 0.335 \\
         \hline
    \end{tabular}
    \label{tab:orbits}
\end{table}

% Astrometric errors
Astrometric errors were assigned either by using the values reported by observers in the ADES astrometric format, when available, or with the astrometric error model by \citet{veres-etal_2017} when unavailable. Star catalogue biases are handled by using the model by \citet{farnocchia-etal_2015b}. Astrometric errors are then adjusted to consider timing errors using the method described in \citet{farnocchia-etal_2022}, which increases the error according to the sky-plane motion at the observation epoch. When the observer does not report the timing error, a default error of 1 s is assumed. Imminent impactors like 2023~CX1 and 2024~BX1 have a large sky-plane motion near the impact epoch, and this causes observations taken right before the impact to be generally de-weighted. However, timing errors mostly affect the along-track component, while the cross-track component is generally better constrained. Therefore, a part of the information is still retained in the orbit determination process when timing errors are considered. On the contrary, these observations would be discarded as outliers without taking into account timing errors, therefore losing all the information.

% Impact corridor computation
After the orbit determination process, the impact location at 100 km height from the surface of the Earth is computed with the semi-linear method by \citet{dimare-etal_2020}. The algorithm provides in output the location of the confidence ellipses at 1-$\sigma$, 3-$\sigma$, and 5-$\sigma$ level, together with the corresponding relative velocity, azimuth, and elevation angles.

\begin{figure*}[!ht]
    \centering
    \includegraphics[width=0.85\textwidth]{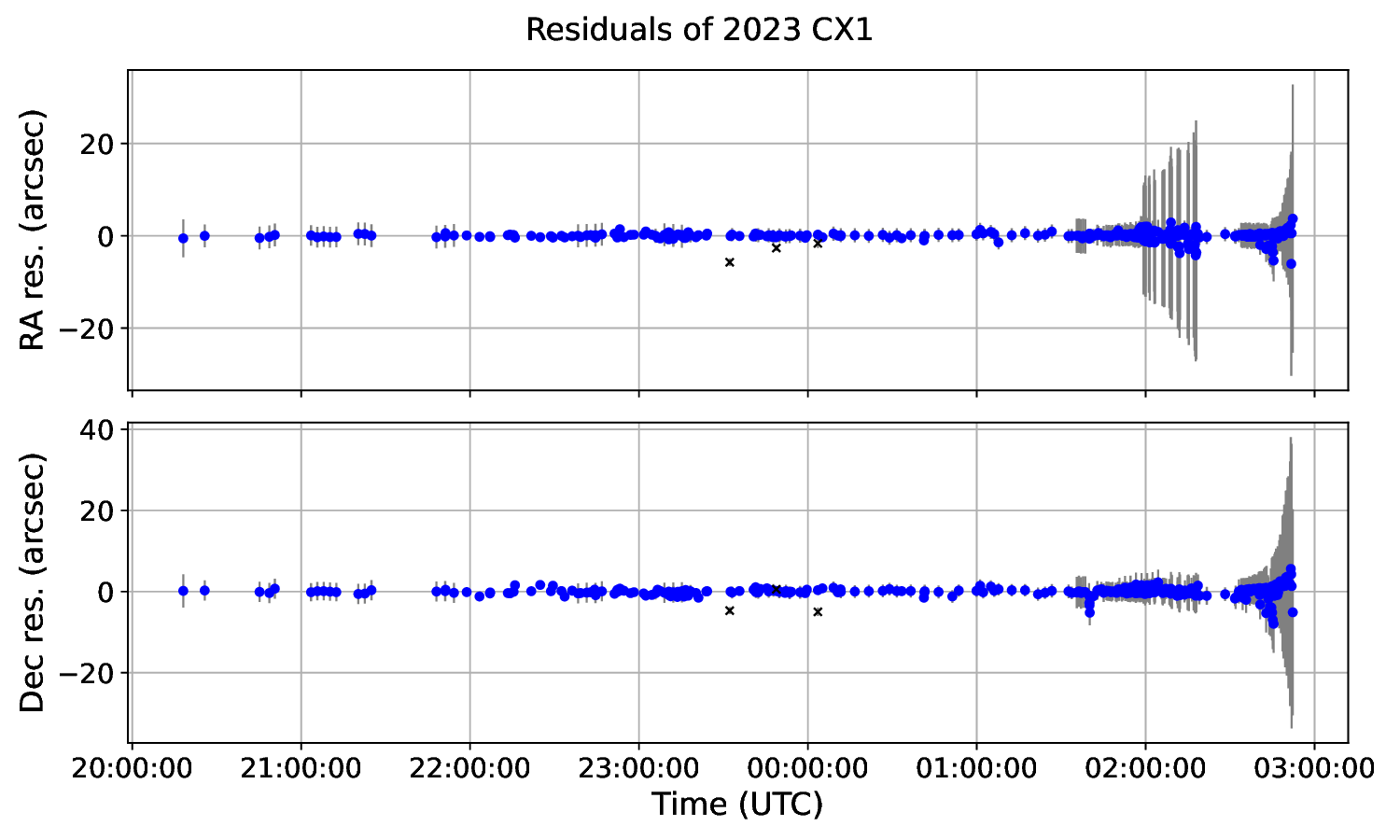}
    \includegraphics[width=0.85\textwidth]{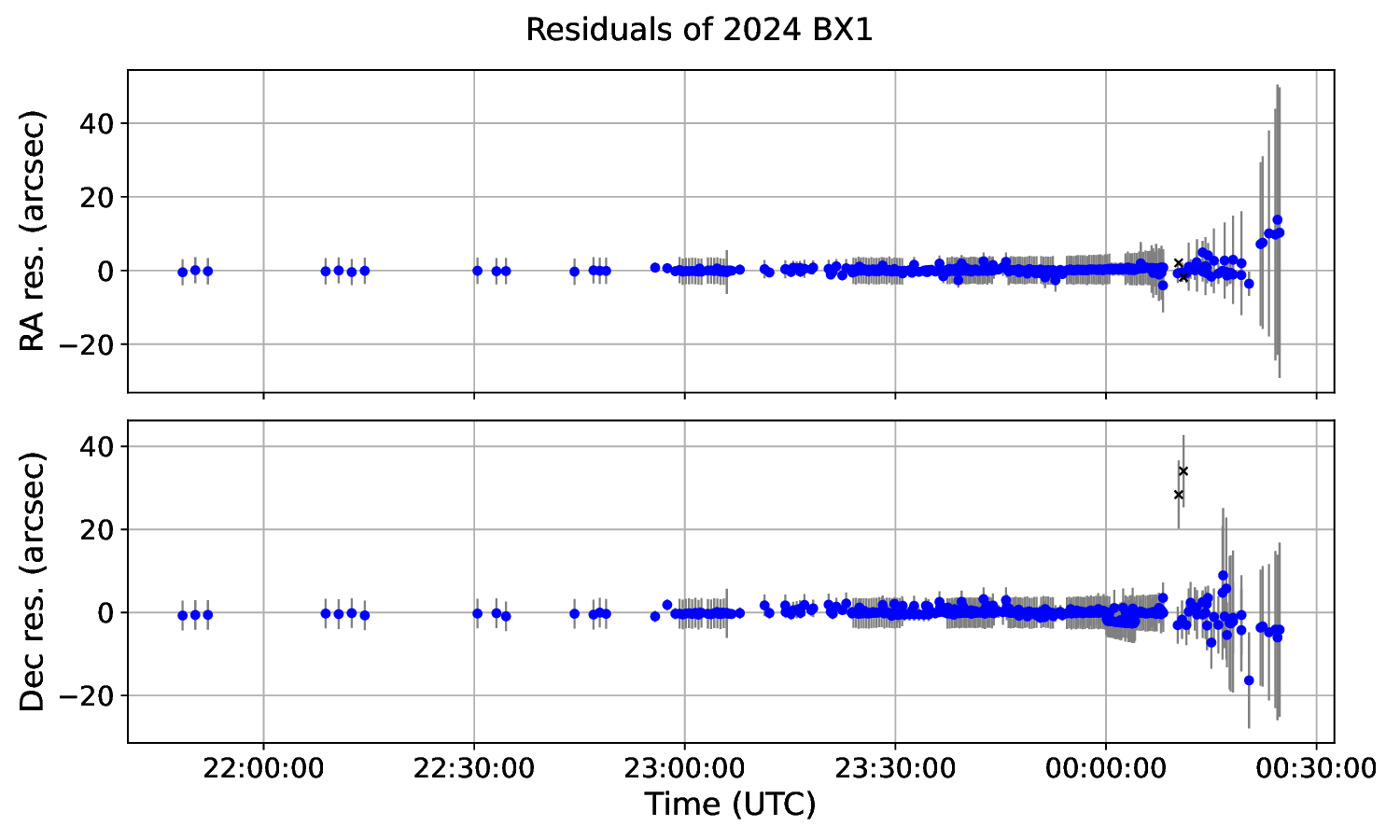}
    \caption{RA and Dec. post-fit residuals of 2023~CX1 (first and second panels) and of 2024~BX1 (third and fourth panels). Blue dots are observations that are accepted for the orbit determination procedure, while black crosses are observations identified as outliers. Error bars correspond to the post-fit astrometric uncertainties assumed for orbit determination.}
    \label{fig:residuals}
\end{figure*}

A total of 368 observations for 2023~CX1 and 323 for 2024~BX1 are available at the time of this work (including additional measurements and remeasurements submitted to the MPC after the events), and the corresponding astrometric records in ADES astrometric format were downloaded from the MPC through the new MPC Explorer\footnote{\url{https://data.minorplanetcenter.net/explorer/}} service. 
Table~\ref{tab:orbits} shows the orbital elements of the two impacting asteroids, together with their normalized root-mean-square (RMS) of the residuals. The reference epoch corresponds to the weighted average of the epochs of the optical observations, as outlined in \citet{milani-gronchi_2009}. Note that since these epochs are close to the impact time, the effects of Earth gravity have already significantly perturbed the motion, and the given orbital elements are valid only at the given epoch. Still, orbital elements are reported with a significant accuracy to allow repeatability of the results.
Astrometric residuals in Right Ascension (RA) and Declination (Dec) with respect to the time are shown in Fig.~\ref{fig:residuals}. Blue dots are observations used in the final orbital fit, while error bars correspond to the post-fit astrometric error computed by considering the proper motion on the sky. Three observations of 2023~CX1 and two of 2024~BX1 were discarded as outliers, as their residuals are way off from the assumed astrometric uncertainties. In both cases, large residuals near the impact epoch are caused by the fast sky-plane motion.  %about XXX for 2023~CX1 and XXX for 2024~BX1.  
Still, these observations are accepted for the orbital fit because they are de-weighted by the effect of timing errors. The final normalized RMS of the residuals was 0.400 for 2023~CX1 and 0.335 for 2024~BX1. 
%

% Image with residuals

% Impact corridors
The impact parameters at 100 km altitude are reported in Table~\ref{tab:longlatunc}, together with their 1-$\sigma$ uncertainties. In both cases, the 1-$\sigma$ ellipse is determined to an outstanding precision. The ellipse semi-major axis for 2023~CX1 is about 96 meters long, while it is only about 64 meters for 2024~BX1. The semi-minor axis is even smaller, about 15 and 6 meters for 2023~CX1 and 2024~BX1, respectively. The passage through the 100 km altitude above the atmosphere is also accurately determined, with a formal error smaller than 0.1 s.   

\begin{table}[!ht]
    \centering
    \caption{Impact parameters at 100 km altitude (WGS84 ellipsoid) from Earth surface of 2023~CX1 and 2024~BX1, with their 1-$\sigma$ uncertainties. The velocity is given with respect to the ground. Elevation is the angle that the velocity vector forms with the ground; the module corresponds to the inclination of the trajectory. The azimuth gives the direction of the velocity vector seen by the observer on Earth, so to get the incoming direction, we must add 180 degrees.}
    \begin{tabular}{ccc}
    \hline
       Object          & 2023~CX1 & 2024~BX1 \\
    \hline
       Time (UTC)                     & 2023-02-13 02:59:13.387 $\pm$ 0.070 s   & 2024-01-21 00:32:37.712 $\pm$ 0.069 s \\
       Latitude (deg)                 & $49.919022   \pm 0.00012$   & $52.584477 \pm 0.000176$ \\
       East Longitude (deg)           & $-0.1467625  \pm 0.0006545$ &$12.356914 \pm 0.000376$ \\
       Velocity (km s$^{-1}$)         & $14.01645    \pm 0.000185$   &$15.197428 \pm 0.000266$ \\
       Elevation (deg)                & $-49.153055  \pm 0.00051$    &$-75.605503 \pm 0.000293$ \\
       Azimuth (deg)                  & $101.386628  \pm 0.000393$  &$73.846168\pm 0.000300$ \\
       1-$\sigma$ semi-major axis (m) & 95.93 & 63.88 \\
       1-$\sigma$ semi-minor axis (m) & 14.23 & 6.12 \\
    \hline
    \end{tabular}
    \label{tab:longlatunc}
\end{table}

% Table with impact parameters

\begin{figure}
    \centering
    \includegraphics[width=0.8\textwidth]{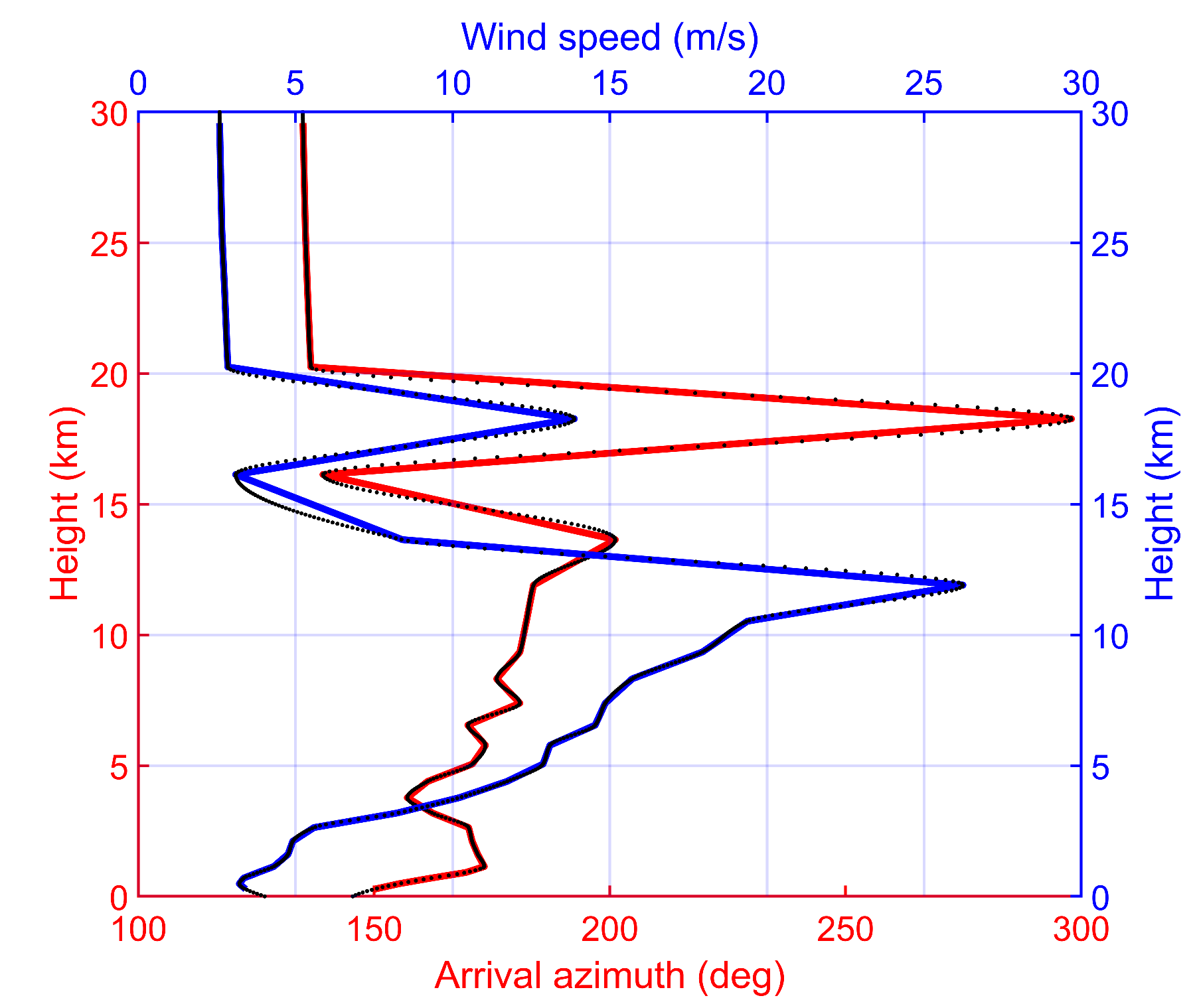}
    \caption{The atmospheric profile used for the 2023 CX1 event, computed by Meteo Expert for 13 February 2023, 03 UTC at Lat. $49.7979^\circ$ N and Long. $0.7533^\circ$ E, a point near the start of the dark flight phase. The modulus of the wind speed (m/s) and the arrival azimuth of the wind (deg) are shown. The dotted line interpolates the data provided by the atmospheric profile using a shape-preserving piecewise cubic interpolation.}
    \label{fig:2023CX1_atmospheric_profile}
\end{figure}

\section{The atmospheric profile}
\label{sec:atmospheric_profile}
A good atmospheric profile is one of the critical points, along with the value of strength, for computing a realistic strewn field. The essential quantities it must report are the air density $\rho_a$ at different heights (or the pressure and temperature from which the density can be calculated) and the speed and direction of the wind $W$. In our profile, the wind components are represented by the standard variables UGRD and VGRD, abbreviated as $U$ and $V$. The $U$ wind component is parallel to the x-axis (i.e. longitude). A positive $U$ wind comes from the west, and a negative $U$ wind comes from the east. The $V$ wind component is parallel to the y-axis (i.e. latitude). All these quantities enter directly into the equations of the fall model used; see Eq.~(\ref{eq:motion_inertial}).\\
However, for our computations, we proceeded step by step. As a first approximation, the atmospheric profile was selected using the radio-sounding data closest to the atmospheric entry point the University of Wyoming provided\footnote{\url{https://weather.uwyo.edu/upperair/europe.html}}. These radio soundings reach about 30 km in height and are made at 00 UTC and 12 UTC. The atmospheric profile from a radio-sounding can be valid for a variable time depending on the atmospheric conditions and the distance between the sounding measurement and the point of interest. In a stable situation, in the middle troposphere, it can be representative of a few hours or even many hours in a variable radius estimated on average between 100 and 200 km. Close to the ground and, more generally, in the boundary layer, it can depend on numerous factors related to diurnal variations and the seasonal period, among other possible causes. It can be valid for 2-3 hours on average in a much smaller radius. In unstable situations, however, the validity in the middle troposphere can be limited to a period varying between 1 and 4 hours; in the boundary layer, it could also be limited to one hour, and the radius can be limited between a few km and a few tens of km. In short, the validity limits of a radio-sounding are difficult to know a priori; for this reason, it must be used with caution, and it is always preferable to use an ad-hoc atmospheric profile. In the cases of the fall of 2023 CX1 and 2024 BX1, we were pretty lucky. In the first case, the weather had a robust high-pressure promontory, with circulation at altitude, while on the ground, it remained the same throughout the day. Also, for 2024 BX1, there was always a high-pressure headland, but less intense. So, the circulatory regime remained the same for several hours in both cases.\\
After computing a first fall model with $S\approx 1$ MPa using this very approximate atmospheric profile, the height and geographical coordinates of the start of the dark flight are obtained (assumed when the speed of the 1 kg final mass fragment drops below 03 km/s). At this point, the atmospheric profile for this location can be computed, and the model can be reiterated. Usually, a recalculation of the profile is not necessary because the difference between the starting points of the two dark flights is only a few km. \\

\begin{figure}
    \centering
    \includegraphics[width=0.8\textwidth]{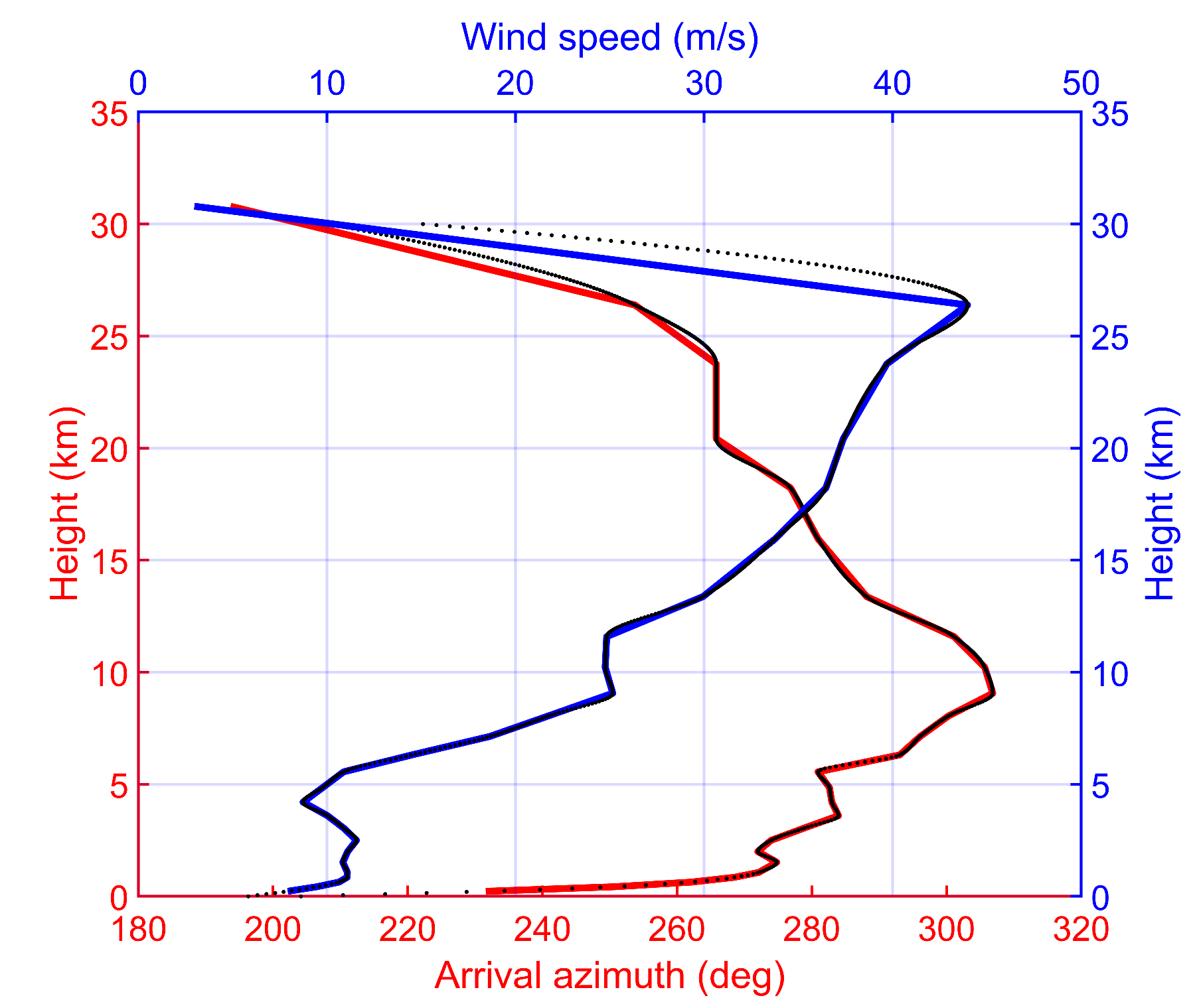}
    \caption{The atmospheric profile used for the 2024 BX1 event, computed by Meteo Expert for 24 January 2024, 00 UTC at Lat. $52.6331^\circ$ N and Long. $12.6379^\circ$ E, a point near the start of the dark flight phase. The modulus of the wind speed (m/s) and the arrival azimuth of the wind (deg) are shown. The dotted line interpolates the data provided by the atmospheric profile using a shape-preserving piecewise cubic interpolation.}
    \label{fig:2024BX1_atmospheric_profile}
\end{figure}

\noindent Meteorological data for our models came from IOIS (Integrated Observations Ingesting System), elaborated and used at Meteo Expert\footnote{\url{https://www.meteo.expert/}}, a private organisation providing meteorological services where weather models are internally developed and applied. Meteo Expert provides paid services, and anyone can request data. Depending on the service, there are costs to bear. In this case, it was a scientific collaboration in agreement with the structure and without charges. Third-party requests may incur a cost to the requester unless other agreements are made. The atmospheric profile is computed by rounding the time to the nearest hour. Its validity is superior to that of a simple radio-sounding because it considers all the available meteorological data and includes evolution over time, given that the data is integrated hourly and localised near the dark flight starting point.\\
This is an important aspect, and we want to underline it: atmospheric data in the troposphere and beyond come from several sources like radiosondes, satellites and planes. All data are integrated, allowing the global surface and three-dimensional atmospheric fields to be obtained. The field's accuracy depends on the spatial and temporal density of the observations. Still, data assimilation, a process of integrating observational data into a numerical model, combines real-world measurements with the model to create an optimised representation of the current state of the atmosphere. This process distributes data in those areas where observations are scarce, allowing us to obtain reliable atmospheric profiles worldwide.
The model that reanalyses all data can provide globally three-dimensional atmospheric fields at hourly intervals on a grid whose horizontal resolution is on the order of tens of km, and they can be mapped to the exact coordinates of the beginning of the dark flight with an appropriate interpolation. \\
An atmospheric profile at a given geographic location can also be forecasted after several hours; so, knowing the small asteroid that will collide with the Earth and the point of entry into the atmosphere at 100 km of altitude, it is possible to predict in advance and with a good approximation, as we will see later, where the strewn field will be located. For this reason, it is possible to speak of a ``strewn field ab initio''.\\

\begin{figure}
    \centering
    \includegraphics[width=0.8\textwidth]{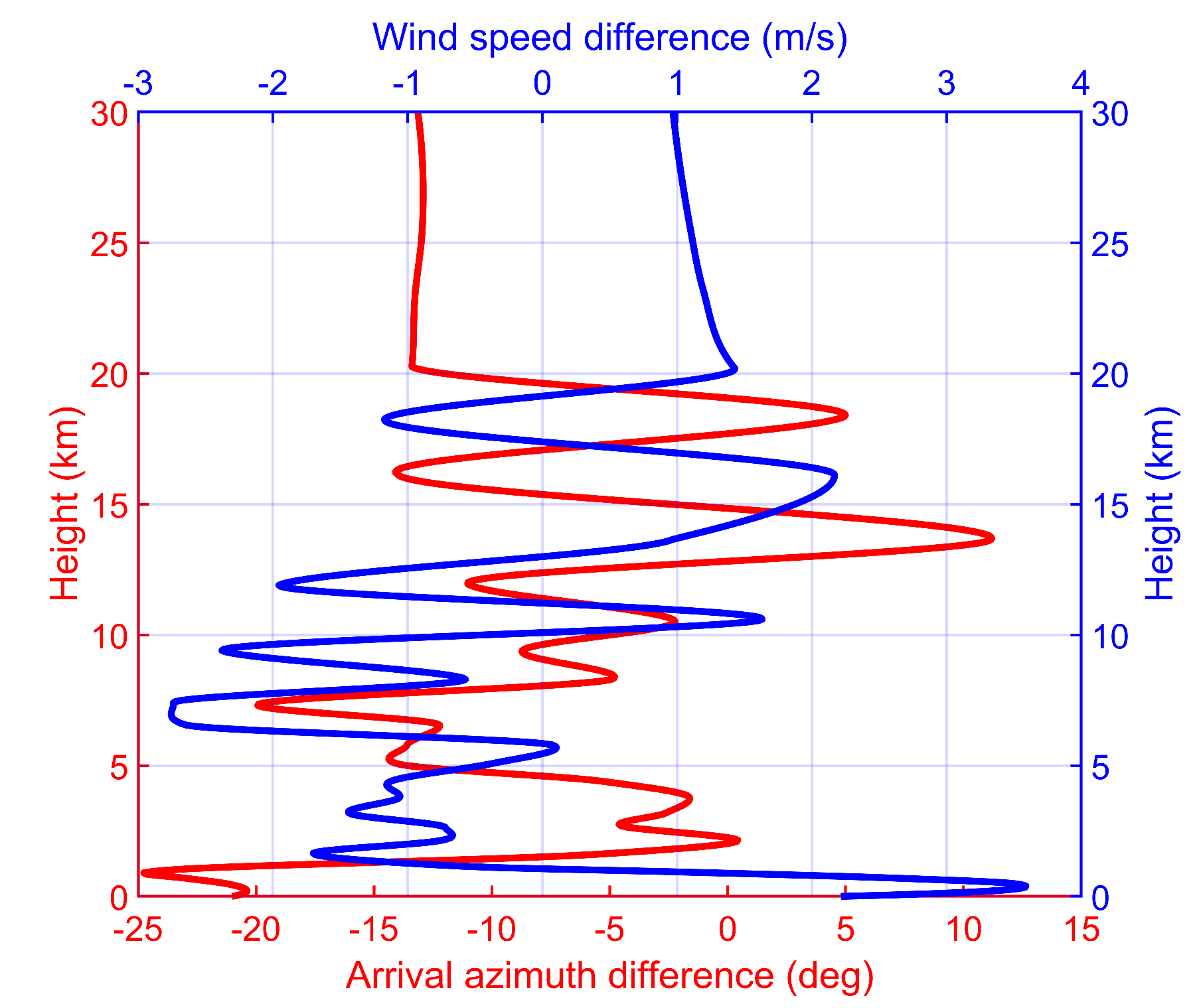}
    \caption{The difference between two atmospheric profiles for the 2023 CX1 event, computed by Meteo Expert for Feb 13, 2023, at 00 UTC and 03 UTC at Lat. $49.7979^\circ$ N and Long. $0.7533^\circ$ E. The difference in speed is in the order of a few m/s, while the difference in arrival azimuth is about $10^\circ$. These differences do not change the strewn field in a relevant way.}
    \label{fig:2023CX1_difference_atmospheric_profile}
\end{figure}

\noindent Considering that the fireball model starts from 100 km in height, as a profile, we use the Meteo Expert density, wind and direction values for heights below 30 km and the 1976 US standard atmosphere model above. This division makes sense because the strongest winds on Earth are jet streams in the upper troposphere. The strongest jet streams are the polar jets around the tropospheric polar vortex, which are large low-pressure areas located over the North Pole (for the northern hemisphere) at an altitude between 5 and 9 km where it can reach speeds up to $100 ~\textrm{m }\textrm{s}^{-1}$. There are also winds in the high stratosphere above 30 km and in the mesosphere (between 50-100 km), but the cosmic speed of the falling asteroid is much higher, so their effect on the body trajectory is negligible \citep{DOWLING2007169}.\\
The computed atmospheric model from Meteo Expert gives the desired quantity for discrete height only, so interpolation is necessary to put the right data in the numerical integration procedure. We use a shape-preserving piecewise cubic interpolation of the values at neighbouring grid points. This algorithm requires at least 4 points and can follow the irregular trend of atmospheric parameters quite well. In Fig.~\ref{fig:2023CX1_atmospheric_profile} and Fig.~\ref{fig:2024BX1_atmospheric_profile}, we have plotted the modulus of the wind speed $W$ with the wind arrival azimuth $AZ$ for 2023 CX1 and 2024 BX1 events; the interpolation result is also shown. The $AZ$ value was computed from the $U$ and $V$ wind component with the transformation equation:

\begin{equation}
AZ = 180^\circ + \arctantwo\left( V, U\right)
\label{eq:wind_azimuth}
\end{equation}

\noindent where the $\arctantwo(Y, X)$ function returns the four-quadrant inverse tangent of Y and X. In this way, it is possible to have an intuitive view of the atmospheric profile that will be useful next to explain the strewn field structure qualitatively. In Fig.~\ref{fig:2023CX1_difference_atmospheric_profile}, we show the difference between two atmospheric profiles for 2023 CX1 computed for the same position at 00 UTC and 03 UTC, the true time of the asteroid fall. As we can see, the difference in the wind speed is a few m/s, while the difference in arrival azimuth is $25^\circ$ at most. This figure provides an example of the change that occurs in the atmospheric profile in a few hours in a quiet meteorological condition and will be useful in estimating the corresponding change in meteorite position. Of course, things can change in rapidly evolving meteorological situations, so it should not always be a valid example.

\section{The 2023 CX1 fall model and the strewn field}
\label{sec:2023CX1_strewn_field}
\begin{figure}
    \centering
    \includegraphics[width=1.0\textwidth]{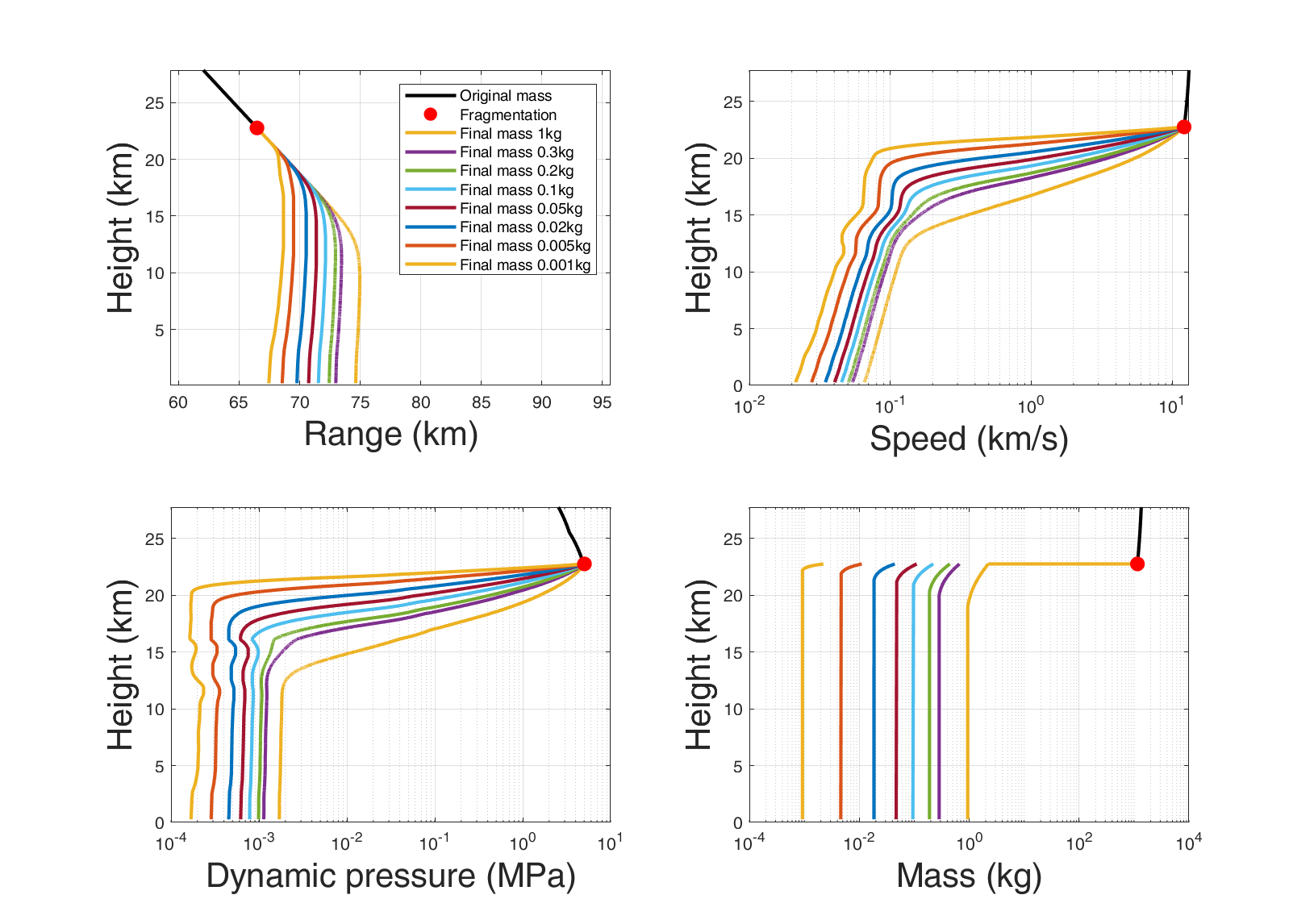}
    \caption{The results of the fragmentation model for the 2023 CX1 fall with $S\approx 5$ MPa and the Meteo Expert atmospheric profile of 03 UTC. The different colours refer to the assumed fragments to sample the strewn field. As expected, the lighter fragments fall before the heaviest in the height vs. range plot. In the height vs. mass plot, note that the fragment masses are subject to a residual ablation after fragmentation. For more readability, some scales in the graphs are logarithmic.}
    \label{fig:2023CX1_model}
\end{figure}

We applied our fragmentation model to the 2023 CX1 event, using the parameters derived from orbit computation and reported in Table~\ref{tab:longlatunc} as starting values. So, the starting height is $h=100$ km above the WGS84 zero reference height at a point with $~\textrm{Lat}_0=49.919022^\circ$ N and $~\textrm{Long}_0=0.1467625^\circ$ W. The starting speed is $v_0=14.01645$ km/s, with a trajectory inclination $I_0=49.153055^\circ$ and the incoming azimuth of $A_0=101.386628^\circ + 180^\circ = 281.386628^\circ$. Remember that elevation from Table~\ref{tab:longlatunc} is the angle that the speed vector forms with the ground; the module corresponds to the inclination of the trajectory. The azimuth, always from Table~\ref{tab:longlatunc}, gives the direction of the speed vector seen by the observer on Earth, so to get the azimuth of the incoming fireball direction, we must add 180 degrees. The uncertainties in the starting values are so small that, for our purposes, they are negligible. We also assume a diameter of about 1 metre and a mean density of about $\rho_m\approx 3000 ~\textrm{kg}~\textrm{m}^{-3}$, so the kinetic energy is about $3.69\cdot 10^{-5}$ Mt. The atmospheric profile was computed for 03 UTC at Lat. $49.7979^\circ$ N and Long. $0.7533^\circ$ E, a point near the start of the dark flight phase (see Fig.~\ref{fig:2023CX1_atmospheric_profile}). As an example of the effects on the strewn field of an atmospheric profile from radio-sounding, we also used the data from the Herstmonceux meteo station (number 3882, Lat. $50.90^\circ$, Long. $0.32^\circ$), at 00 UTC of 13 February 2024, located in the UK near Hastings, about 125 km north of the real strewn field.\\
According to our model with $S\approx 5$ MPa, fragmentation occurred approximately 7.4 s after the start, and the fireball phase for the 1 kg fragment ended after about 0.9 s, while the dark flight started at a height of about 19 km. Finally, the fragment touches the ground at about 240 km/h; see Fig.~\ref{fig:2023CX1_model} for a plot of the model. \\

\begin{figure}
    \centering
    \includegraphics[width=1.0\textwidth]{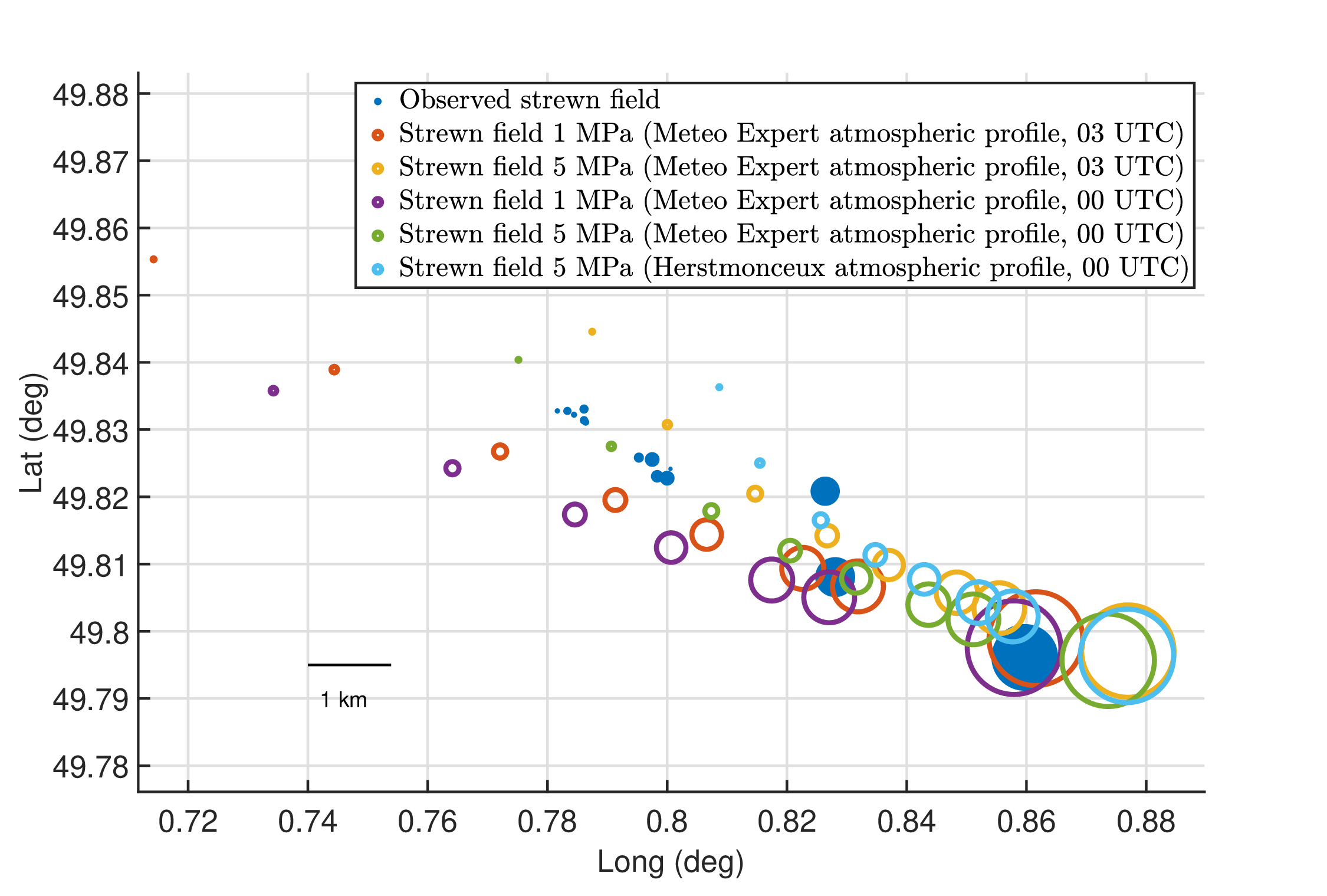}
    \caption{A comparison between the observed strewn field of 2023 CX1 fall from Table~\ref{tab:meteorites_2023CX1} and the theoretical ab initio strewn field with mean strengths of 1-5 MPa and the Meteo Expert atmospheric profile. The circles are proportional to the mass of the recovered meteorite, and as a comparison, the strewn field with the Herstmonceux data is also present. As we can see, the fit is better for $S\approx 5$ MPa with the Meteo Expert atmospheric profile,  see Table~\ref{tab:rms1} for a numerical comparison. Coincidentally, the atmospheric profile with strength 5 MPa at 00 UTC provides slightly better results than that at 03 UTC.}
    \label{fig:2023CX1_strewn_field}
\end{figure}

\noindent If, in the fall model, we used an average strength equal to $S\approx 1$ MPa, the fragmentation is at a height of about 34 km from the ground, while the dark flight for a 1 kg final mass started at a height of about 22 km, and the fragment touches the ground at about 237 km/h. Considering that under the height of 10 km, where the air density is highest, the azimuth of the wind was around $160^\circ$, forming an angle of about $120^\circ$ with the fireball's arrival azimuth (see Fig.~\ref{fig:2023CX1_atmospheric_profile}), the fragments were moving partially against the wind. So, in this condition, the smaller fragments tend to be carried away in the same direction as the wind. Furthermore, if fragmentation occurs higher (as for $S\approx 1$ MPa case), the fragments reach the ground at a slower speed, and this is the reason why the strewn field with $S\approx 1$ MPa is shifted westward by a few km with respect to the strewn field with $S\approx 5$ MPa, see Fig.~\ref{fig:2023CX1_strewn_field}. Interestingly, the position difference between the ab initio strewn fields is lower for the major fragment. This is due to the fragment's mass: the bigger fragments are less sensible to the wind direction, so it is more probable to recover big meteorites ($\sim 1 ~\textrm{kg}$), even with the approximations of an ab initio strewn field. The two theoretical strewn fields, the first with $S\approx 1$ MPa and the second with $S\approx 5$ MPa, form a ``V-shaped'' structure that, excluding the uncertainty due to atmospheric profile, delimits the most probable region where to search for meteorites.  \\

\section{The 2024 BX1 fall model and the strewn field}
\label{sec:2024BX1_strewn_field}
Also, in the case of 2024 BX1, the procedure was the same as for 2023 CX1. The initial conditions upon entry into the atmosphere are the following (see Table~\ref{tab:longlatunc}): $h=100$ km, latitude $~\textrm{Lat}_0=52.584477^\circ$ N, longitude $~\textrm{Long}_0=12.356914^\circ$ E, $v_0=15.197428$ km/s, $I_0=75.605503^\circ$ and $A_0=253.846168^\circ$. To simulate a computation immediately after a fall, as the starting diameter, we had assumed 1 m, with a mean density of $\rho_m\approx 3000 ~\textrm{kg}~\textrm{m}^{-3}$; so we ignore the 0.44 m diameter estimate from the European Fireball network because we want to simulate what would have been obtained with only the available orbital data. 

\begin{figure}
    \centering
    \includegraphics[width=1.0\textwidth]{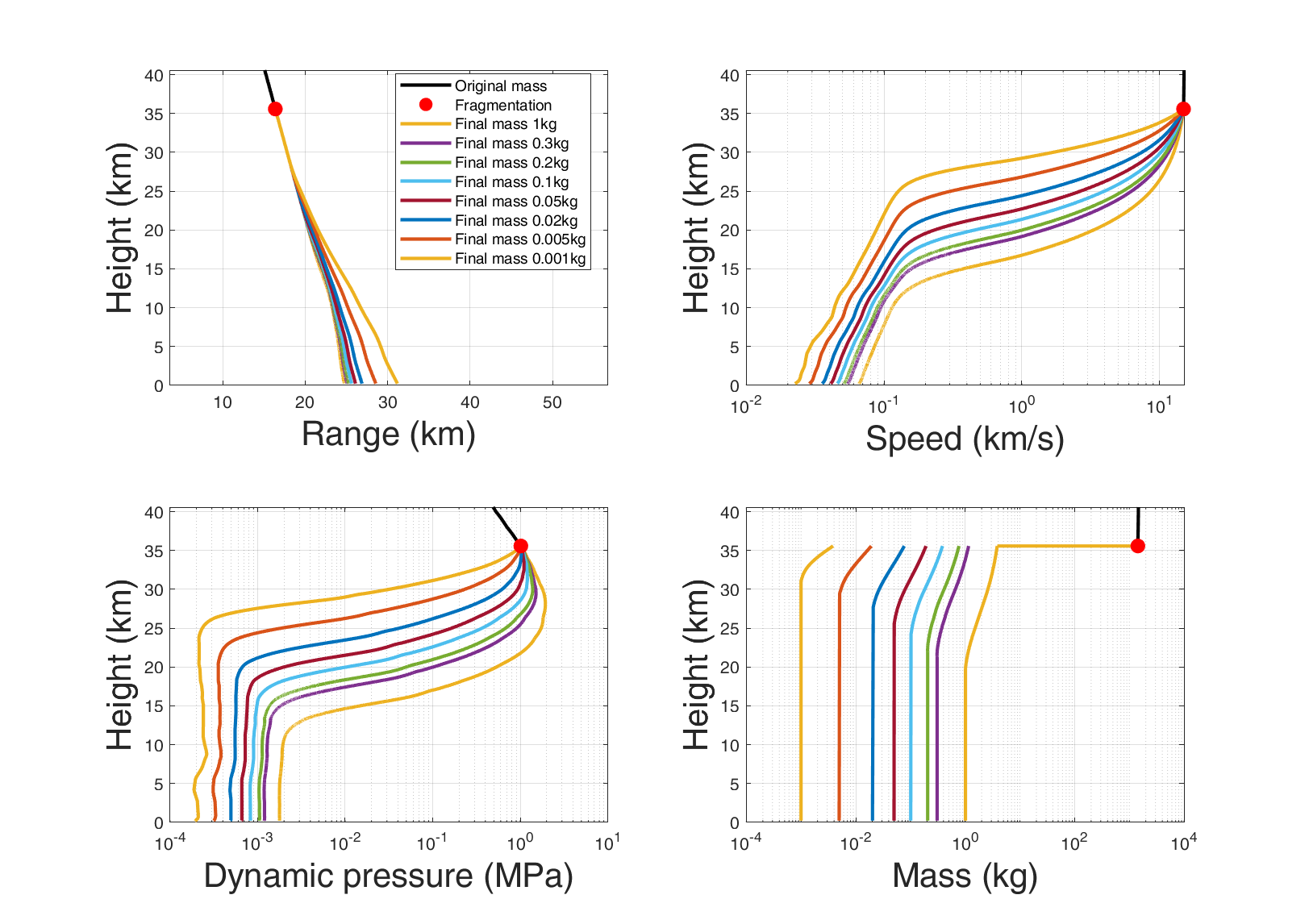}
    \caption{The results of the fragmentation model for the 2024~BX1 fall with $S\approx 1$ MPa and the Meteo Expert atmospheric profile. The different colours refer to the assumed fragments to sample the strewn field. Note that in the height vs. range plot, the sequence of the fragment's mass is reversed with respect to 2023~CX1 because of the wind. In the height vs. mass plot, note that the fragment masses are subject to a residual ablation after fragmentation. For more readability, some scales in the graphs are logarithmic.}
    
    \label{fig:2024BX1_model}
\end{figure}

\begin{figure}
    \centering
    \includegraphics[width=1.0\textwidth]{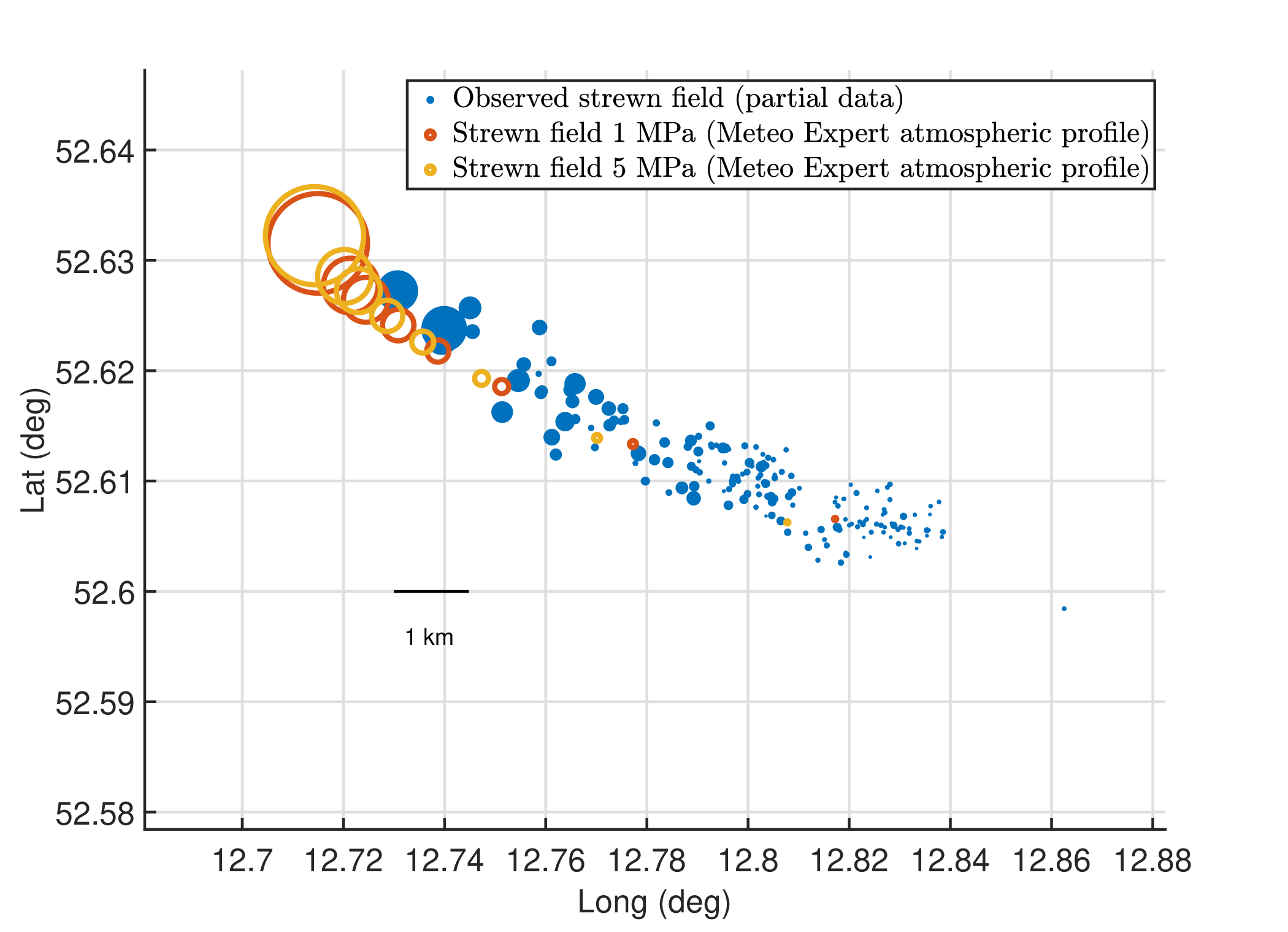}
    \caption{A comparison between the observed strewn field of 2024 BX1 fall from \cite{Bischoff2024} and the two theoretical ab initio strewn fields with a mean strength of 1 and 5 MPa. This figure shows the positions of the 185 meteorites of 202 recovered, whose positions are known. The circles are proportional to the mass of the recovered meteorite, and the agreement is slightly better for the strewn field with $S\approx 1$ MPa; see Table~\ref{tab:rms1} for a numerical comparison.}
    \label{fig:2024BX1_strewn_field}
\end{figure}

\noindent In the scenario with $S\approx 1$ MPa, the fragmentation occurred about 4.4 s after the starting time at a height of about 35 km, and the fireball phase ended after 6.3 s. The duration of the fireball phase is in agreement with the observed light curves; see Section~\ref{sec:2023CX1_2024BX1_event}. Most recorded fragmentations from all-sky cameras also fall between 30 and 38 km altitude, with aerodynamic pressure in the 1-2 MPa range \citep{Spurny2024}. In the scenario with $S\approx 5$ MPa, the fragmentation occurs later (about 5.2 s, height 24 km), and the fireball phase ends after 5.5 s. Fig.~\ref{fig:2024BX1_model} reported the model results for the $S\approx 1$ MPa scenario, which appears more adequate for the event description because the main fragmentation height is near the observed one, see Table~\ref{tab:rms1} for a quantitative comparison between the strewn fields. In Fig.~\ref{fig:2024BX1_strewn_field}, there is a comparison between the ab initio strewn fields and the observed one. Qualitatively, the observed strewn field has a uniform width of about 1 km, with a length of about 9-10 km: the larger meteorites are in the west, while the smaller ones extend towards the east, as expected. The computed strewn fields are very near the observed ones.\\
The overlap of the theoretical strewn fields results from the high inclination of the fireball trajectory, which tends to make the larger fragments fall in the same place, regardless of the fragmentation height determined by the $S$ value. The wind arrival azimuth influenced the position of the smaller fragments: under the height of 10 km where the wind determines the fragment's path, the azimuth from which the wind arrives is approximately $280^\circ$ (see Fig.~\ref{fig:2024BX1_atmospheric_profile}) so, the theoretical strewn field tends to be extended in the southeast direction.

\section{The 2008~TC3 fall model and the strewn field}
\label{sec:2008TC3}
 As a further test for our simple fall model with a real wind profile, we analyse the well-known 2008~TC3 case, the first asteroid discovered a few hours before Earth's impact \citep[]{shaddad-etal_2010, Farnocchia2017}. This is a case where there were no triangulations of the fireball from the ground but only satellite observations \citep{Borovicka2009} and eyewitness accounts, so it falls into the category of events where the application of our model is useful for identifying the strewn field with orbital data. \\

% Orbital elements table
\begin{table}[!ht]
    \centering
    \caption{Keplerian orbital elements of 2008~TC3, with the corresponding epoch, expressed in MJD, and impact parameters at 100 km altitude from Earth's surface. Errors refer to the 1-$\sigma$ formal uncertainties.}
    \begin{tabular}{cccc}
         \hline
      Orbital Data                 & Value                            & Impact Parameter               &  Value \\
         \hline
      Epoch (MJD)                  & 54745.8110 TDT                    & Time (UTC)                     & 2008-10-07 02:45:30.09 $\pm$ 0.14 s       \\ 
      Semi-major axis (au)         &  $   1.284115   \pm  0.000011 $   & Latitude (deg)                 & $21.0884  \pm 0.0009$       \\ 
      Eccentricity                 &  $   0.294852   \pm  0.000007 $   & East Longitude (deg)           & $30.5347 \pm 0.0038$        \\ 
      Inclination (deg)            &  $   2.403189     \pm  0.000057 $     & Velocity (km s$^{-1}$)         & $12.38041  \pm 0.00005$       \\ 
      Longitude of node (deg)      &  $  194.11280    \pm  0.00000282 $ & Elevation (deg)                & $-20.8360 \pm 0.0030$       \\ 
      Argument of perihelion (deg) &  $  234.0469348    \pm  0.000087 $     & Azimuth (deg)                  & $101.0953  \pm 0.0015$       \\ 
      Mean anomaly  (deg)          &  $ 329.66890    \pm  0.00052 $    & 1-$\sigma$ semi-major axis (m) & 610  \\ 
      Normalized RMS               & 0.230                             & 1-$\sigma$ semi-minor axis (m) & 98   \\
         \hline
    \end{tabular}
    \label{tab:orbit2008TC3}
\end{table}

% 1.2841150331448696E+00  2.9485205833830919E-01   2.4031893392819 194.1128066255609 234.0469348217665  3.2966890153007540E+02
% ! RMS    1.11181E-05   7.63885E-06   5.74784E-05   2.81973E-06   8.76002E-05   5.22798E-04

\begin{figure}
    \centering
    \includegraphics[width=1.0\textwidth]{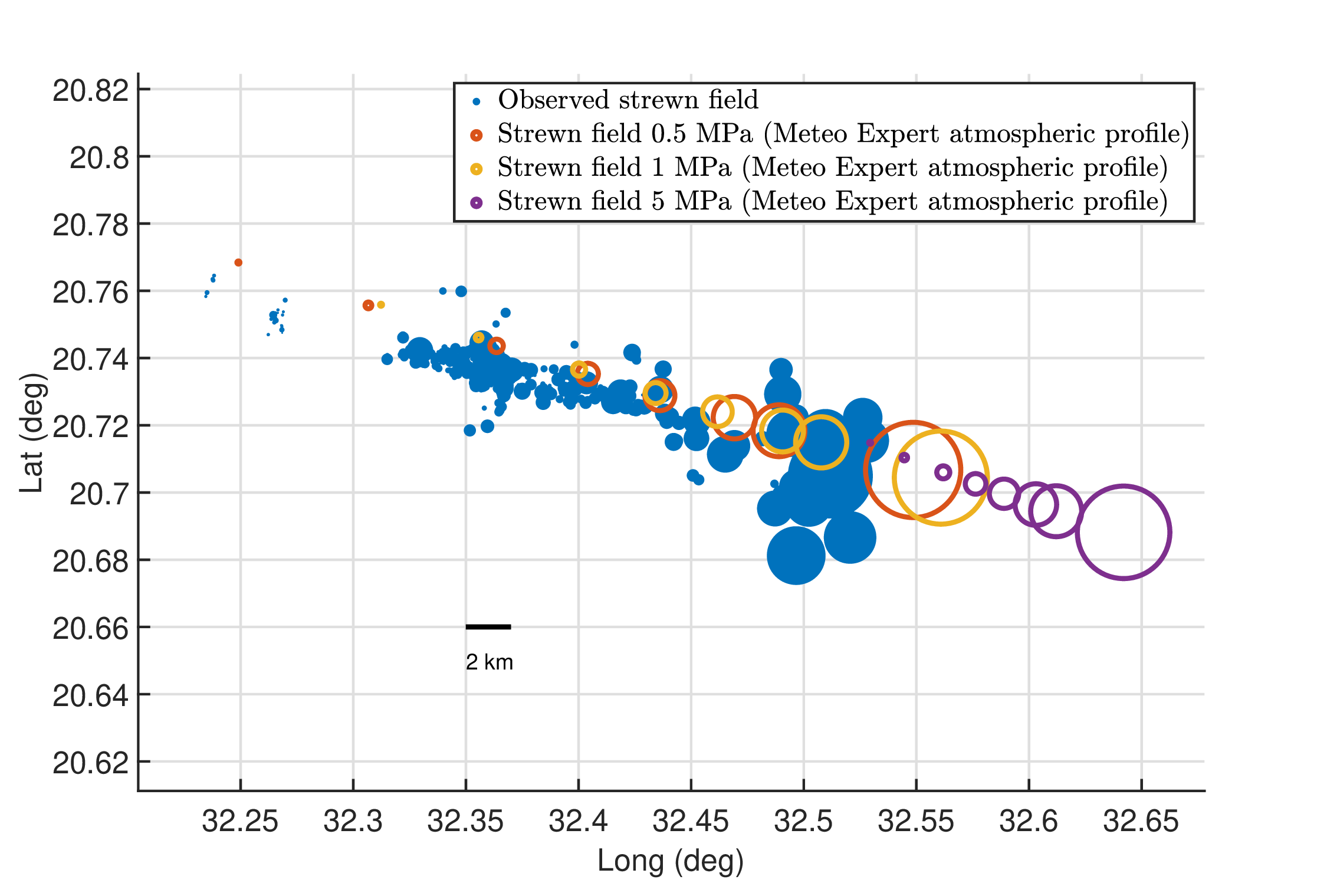}
    \caption{A comparison between the observed strewn field of 2008~TC3 with 624 meteorites of known position and the three theoretical ab initio strewn fields with a mean strength of 0.5, 1 and 5 MPa. The circles are proportional to the mass of the recovered meteorite; as we can see, the agreement is qualitatively better for the strewn field with $S\approx 0.5$ MPa, but also for $S\approx 1$ MPa, the result appears good, see Table~\ref{tab:rms1} for a numerical comparison.}
    \label{fig:2008TC3_strewn_field}
\end{figure}

\begin{figure}
    \centering
    \includegraphics[width=0.8\textwidth]{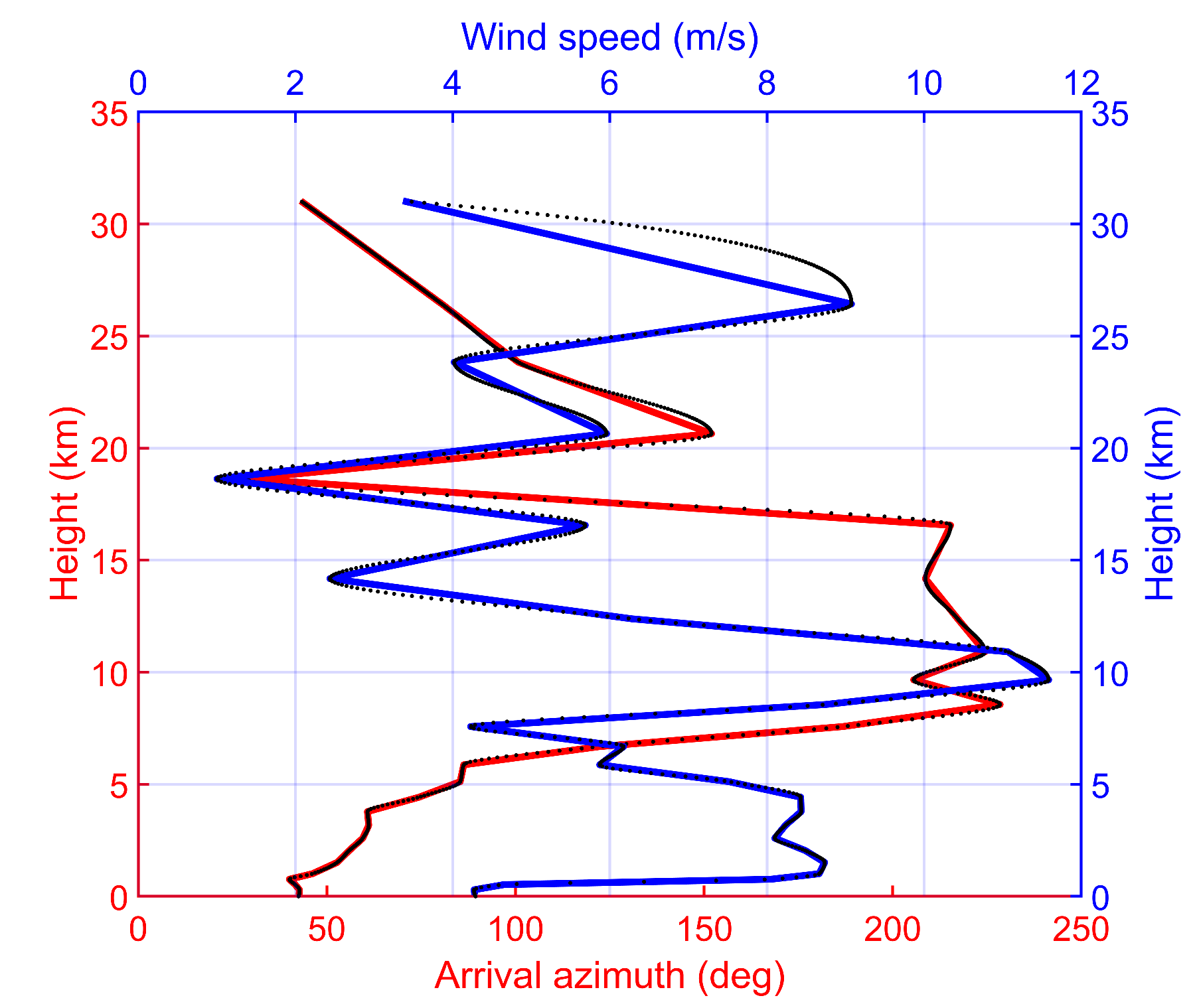}
    \caption{The atmospheric profile used for the 2008 TC3 event, computed by Meteo Expert for 7 October 2008, 03 UTC at Lat. $20.97^\circ$ N and Long. $32.16^\circ$ E. As in the previous atmospheric profile, the modulus of the wind speed (m/s) and the arrival azimuth of the wind (deg) are shown. The dotted line interpolates the data provided by the atmospheric profile using a shape-preserving piecewise cubic interpolation.}
    \label{fig:2008TC3_atmospheric_profile}
\end{figure}

\noindent From satellite data, two flares are present, the first at a height of about 45 km and the main at about 37 km, so at 45 km, the asteroid began to fragment, culminating at 37 km with complete fragmentation. Table~\ref{tab:orbit2008TC3} reports the orbital elements and impact parameters computed with the astrometric observations available at the MPC. Our impact parameters for atmospheric entry, starting height $h=100$ km, $~\textrm{Lat}_0=21.0884^\circ$ N and $~\textrm{Long}_0=30.5347^\circ$ E, starting speed $v_0=12.3804$ km s$^{-1}$, trajectory inclination $I_0=20.8360^\circ$ and incoming azimuth $A_0=101.0953^\circ + 180^\circ = 281.0953^\circ$, agree very well with the values reported by \cite{Farnocchia2017} and \cite{Jenniskens2022}. The absolute magnitude estimated from the astrometric measurements is 30.3; assuming a typical geometric albedo of $p_v\approx 0.1$, we get a diameter of about 3.6 m, a value coherent with Table~\ref{tab:small_impact}. The peculiarities of this fall were the low inclination of the trajectory with respect to the Earth's surface, which increased the observed length of the strewn field to about 30 km, and the fact that the collected meteorites are of different types. About 39 kg of fragments were collected, named Almahata Sitta; most meteorites are classified as anomalous polymict ureilite, but there are also ordinary, enstatite and carbonaceous chondrites \citep{Jenniskens2022}. See \cite{shaddad-etal_2010} for a complete meteorites list, from which we obtained mass and geographical coordinates (WGS84).\\
We adopted the same values as before for the possible meteorite mass that ``samples'' the strewn field, and the comparison between observed and computed strewn fields is reported in Fig~\ref{fig:2008TC3_strewn_field}. The Meteo Expert atmospheric profile was computed for 03 UTC for a point at coordinates Lat. $20.97^\circ$ N and Long. $32.16^\circ$ E, using, as a first guess for dark flight starting point, the atmospheric profile of Jeddah (King Abdul Aziz) collected at 00 UTC on 7 October 2008. Our atmospheric profile shows the presence of weak winds, with a maximum speed of almost 12 m/s, around 10 km height and a direction from southwest to northeast. For altitudes below 5 km, the speed decreases to about 08 m/s and reverses with an arrival direction from the northeast; see Fig~\ref{fig:2008TC3_atmospheric_profile}. These aspects of the wind profile are consistent with \cite{Farnocchia2017}. Overall, the wind did not have a large effect on the position of the meteorites, and this explains why the position of the drag-free ground track field given by \cite{Farnocchia2017}, using orbital elements to predict the position of 2008 TC3 a the 100 km entry point, is in good agreement with the real track position.\\
As we can see, the best-strewn field is that with $S\approx 0.5$ MPa, but also $S\approx 1$ MPa is acceptable for meteorites search, see Fig~\ref{fig:2008TC3_strewn_field}. Instead, the theoretical strewn field with $S\approx 5$ MPa is far from the observed position, even if it is aligned with the previous ones.

\noindent Interestingly, there is a change in cross distributions along the real strewn field line, with the lower masses being more compact than the larger ones; see Fig~\ref{fig:2008TC3_strewn_field}. This change in cross distributions is not due to the wind but to the asteroid's fragmentation process, with the smaller fragments, emitted during the flare at about 45 km height, being held more compact due to the wake vacuum region that the falling asteroid left behind, as demonstrated using a hydrocode simulation by \cite{Jenniskens2022}. In the main flare at 37 km height, the major fragments were released with high relative speeds, resulting in an increased cross-distribution.\\
In our model with $S\approx 0.5$ MPa, fragmentation occurs at a height of about 37 km, approximately 14 s after time zero. This value of the main fragmentation height agrees with satellite observations \citep{Borovicka2009, shaddad-etal_2010}. The fragmentation height drops slightly to 33 km for the model with a strength of 1 MPa and is drastically reduced to 20 km in the case of 5 MPa, causing the notable shift of the strewn field. The good agreement between the observed and simulated strewn fields with $S\approx 0.5$ MPa and only one fragmentation point is probably due to the fact that the smaller fragments were retained in the meteoroid's wake vacuum region until the main fragmentation, behaving as if a single effective fragmentation had occurred. A further example of aligned strewn fields occurs in the case of the fall of 2022~WJ1, which is covered in Appendix~\ref{sec:appendix}.

\section{Discussion}
\label{sec:discussion}
This section will discuss whether strewn fields computed with strengths in the range $S\approx 1 - 5$ MPa are adequate for searching for meteorites on the ground from known asteroid falls without triangulation. First, let's compare the fragments' computed nominal position with that of the recovered meteorites of similar mass, a comparison we made using Google Earth software. In this way, estimating the uncertainty order of magnitude for the ground search will be possible. We start from the fall of 2023 CX1; for the numbering of the recovered fragments and positions, refer to Table~\ref{tab:meteorites_2023CX1} and Fig.~\ref{fig:2023CX1_strewn_field}. The values in the text refer to the Meteo Expert atmospheric profile of 03 UTC, while those in brackets refer to the atmospheric profile of 00 UTC. The meteorite of the theoretical strewn field with $S\approx 5$ MPa and a mass of 0.005 kg is at a distance between 1000 m and 1300 m (600 m and 900 m) from the smaller meteorites truly recovered, the number [2], [3], [4], [5], [6] and [7]. The theoretical meteorite with a mass of 0.02 kg is between 1000 m and 1500 m (800 m and 1200 m) from real fragments of comparable mass ([8], [9], [10] and [11]). Fragment number [13], with a mass of 0.175 kg, is about 1400 m (1200 m) from the predicted position for a 0.2 kg mass. Finally, the biggest recovered meteorite (fragment [14]) is about 1200 m (900 m) from the position of the 1 kg mass. An ``out of the ordinary'' meteorite whose position is approximately 1000 m away from the symmetry axis of the observed strewn field is the fragment [1], approximately 1500 m (1500 m) from the expected position of a 0.1 kg mass. This position anomaly is unlikely due to the lateral expansion speed given by Eq.~(\ref{eq:dispersal_speed}). Not including atmospheric drag, the deviation from the original trajectory would be approximately $0.5^\circ$, but considering that - in this case - the range distance between fragmentation and the fall is about 4.5 km, the deviation would have been, at most, approximately 50 m, too small to justify the observed deviation. It is probably a lift effect that is not included in our model. Overall, the computed$-$observed position ranges from a minimum of 1000 (600 m) to a maximum of 1500 m (1500 m), so we are well within the exploration capabilities of a group of people walking around the nominal impact points. In this case, the atmospheric profile computed for 00 UTC gave slightly better results than 03 UTC, even if the opposite could be expected. The differences between the positions observed and computed for the two profiles show that the differences due to a slight wind regime change are a few hundred meters. \\
When considering the model with $S\approx 1$ MPa with 03 UTC atmospheric profile, things get worse for little masses and slightly better for large masses. In this case, the distance between 0.005 kg and the recovered meteorites with similar masses rises to about 3.0 km; the distance from 0.02 kg is about 1.9 km; the distance from 0.1 kg is 1.7 km; from 0.2 kg is 400 m and finally from 1 kg is 250 m. Therefore, choosing the right strength $S$ is important to minimise the recovery times, especially when the two strewn fields we assumed as extremes do not overlap but have a ``V'' structure. In this case, it is better to concentrate on the largest meteorite search because they tend to fall close to each other, but remember that we do not know if bigger meteorites exist without fireball observation; this is a working hypothesis of our method because the meteorite masses are free parameters.\\
To estimate more quantitatively the ``goodness'' of the strewn field, we have computed the least squares best-fit curve of the theoretical strewn field (from maximum to minimum sample mass) and calculated its root mean square (RMS) with the observed positions of the meteorites, regardless of the mass. From a physical point of view, it is an estimate of the average distance that must be travelled to go from the theoretical line of the strewn field to any meteorite found. We used a second-order polynomial for the best fit of the theoretical strewn field. We found that the best approximation, with RMS=0.29 km, is the one with $S\approx5$ MPa computed for 00 UTC, followed by the two computed at 03 UTC, all from the Meteo Expert profile. Using the Herstmonceux weather profile ($S\approx5$ MPa for 00 UTC) results in RMS=1.4 km; see Table~\ref{tab:rms1}. As we can see, the atmospheric profile computed for a point near the starting point of the dark flight can lower the meteorites' search distance.

\begin{figure}
    \centering
    \includegraphics[width=0.8\textwidth]{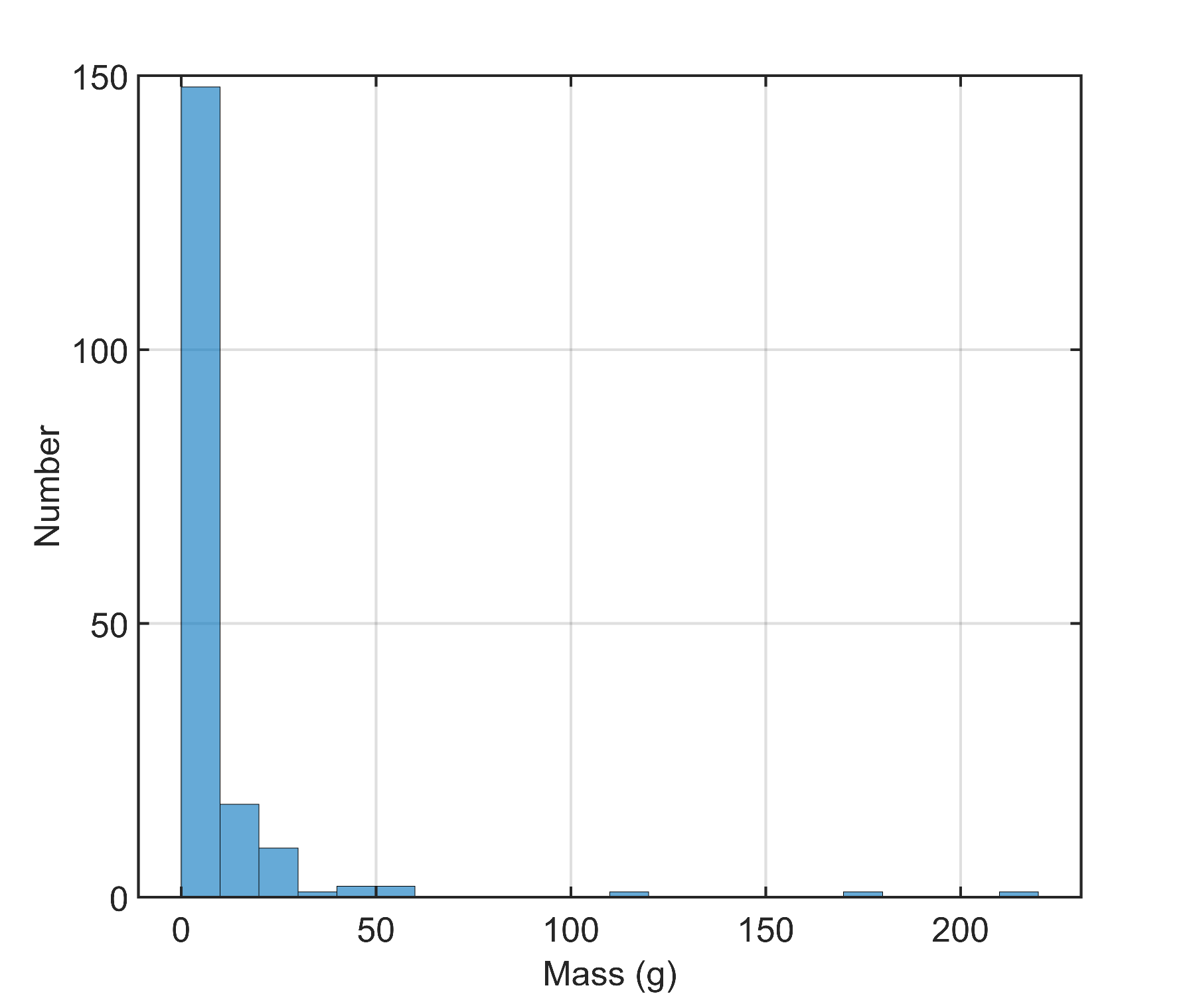}
    \caption{The mass distribution of meteorites with known positions collected in the strewn field of 2024 BX1 \citep{Bischoff2024}. Of the 185 meteorites in total, 151 have a mass of less than 10 g.}
    \label{fig:2024BX1_masses_distribution}
\end{figure}

\noindent About 2024 BX1, the two ab initio strewn fields largely overlap, so the ground search appears less problematic than in the case of 2023 CX1. Limiting ourselves to the largest meteorites, for $S\approx 1$ MPa, the 0.2 kg theoretical position is about 1000 m from Ribbeck [14], while for the $S\approx 5$ MPa case the distance rises to 1200 m. The distance from Ribbeck [1] is approximately 500 m in both cases. Concerning the 0.1 kg final mass, with respect to Ribbeck [2], the distance ranges from 750 m for $S\approx 5$ MPa to 600 m for $S\approx 1$ MPa. The RMS computation tells us that the strewn field computed with $S\approx 1$ MPa is slightly better than the case with $S\approx5$ MPa, see Table~\ref{tab:rms1}. It can be noted that the values of the observed-computed distance for the meteorites with greater mass are lower than the average value of the RMS. This is because most of the meteorites collected have a mass lower than 10 g, see Fig.~\ref{fig:2024BX1_masses_distribution}, which undergo greater dispersion due to their extreme sensitivity to the wind. \\
From these two falls, despite the approximations used in the fragmentation model, which considers a single fragmentation followed by a release of fragments of different mass, all with the same speed and direction with no lateral speed component, the results of the computed strewn field from orbital elements appear pretty good. Overall, the RMS values between the two cases of 2023 CX1 and 2024 BX1 are comparable. In both cases, with the information given by the ab initio strewn fields, the recovery of the meteorites would have been possible because the observed-computed differences are of the order of about 1000 m. Knowledge of the atmospheric profile is fundamental, particularly the prevailing wind regime in the troposphere, which ultimately characterises the position of the meteorites on the ground, especially the smaller ones.\\ 
Regarding 2008~TC3, given the large number of meteorites collected, it's not very practical to make a one-to-one comparison between the positions of theoretical and observed masses, as in the case of 2024~BX1 and 2023~CX1, so we compute the collective RMS of the observed meteorites with the best-fit curve of the theoretical strewn field for the 0.5, 1 and 5 MPa cases, see Table~\ref{tab:rms1}. The RMS values for the cases with $S=0.5$ and 1 MPA are in the range of 1-2 km, a very good value considering the low inclination of the asteroid trajectory with respect to the Earth's surface. In the case of searching for meteorites with only the information available from the heliocentric orbit before the impact and the results of the models, one would have arrived at the actual strewn field even if, starting from the model positions with $S\approx 5$ MPa, it would have been necessary to walk about ten kilometres before finding the fragments. \\
A weak point of our procedure is the lack of a priori knowledge of the asteroid's strength, which is a very important parameter because it determines the height of the fragmentation, so the length of the fragment's path before reaching the ground. This forces us to adopt a range of reasonable values, which may be good for most cases but may not always be valid. The value of the other parameters adopted for the model and listed in Section~\ref{sec:summary_list} are less critical than strength because they determine the ablation rate and the drag. The asteroid's diameter can be estimated from telescope observations, so it is a secondary factor for the position of the strewn field. Overall, the fixed parameter values adopted in the text are generic enough to describe most falls. Based on the results obtained, the number of fragmentations governed by the Weibull modulus value does not appear to be a determining factor for the position of the strewn field; the important thing is that there is a main fragmentation. This behaviour, as we saw in the case of 2008~TC3, is probably caused by the asteroid's wake vacuum region, which retains the smaller fragments that break off in the secondary fragmentations until reaching the main fragmentation of the body when the containment effect ceases, so all the fragments are free to fall independently. In the case of high inclination trajectory, the strewn fields with different strengths are approximately overlapped, and the area to carry out the searches is reduced significantly, while in the low inclination cases, the search area grows consequently. As regards the presence of a possible fast pancake phase in the fragmentation of small asteroids of the order of a meter, based on the results shown in Table~\ref{tab:rms1}, the answer is probably not because the fragmentation alone has lower RMS, even if only slightly, than the pancake model. Probably, the local small-scale turbulences act quickly in dispersing the mass of small fragments of the asteroid, eliminating the pancake phase and making the macroscopic fragments independent from each other. So, the pancake phase appears unnecessary in the case of a small asteroid fall to estimate the possible strewn field position. 

\begin{table}[!ht]
    \centering
    \caption{The RMS values for 2023~CX1, 2024~BX1 and 2008~TC3, computed between the best-fit curve of the theoretical strewn field, characterised by the mean strength $S$ with the atmospheric profile, and the observed positions of the meteorites, regardless of mass. ME=Meteo Expert, H=Herstmonceux. For 2023~CX1, the agreement observed-computed strewn field is better for $S\approx 5$ MPa; for 2024~BX1, the best is for $S\approx 1$ MPa, while for 2008~TC3 is better for $S\approx 0.5$ MPa. The values in brackets refer to the pancake model. As we can see, there is no significant difference between the fragmentation and pancake models. However, the fragmentation model values are slightly lower or equal, except for two values in the 2023~CX1 column referring to the ME atmospheric profile at 00 and 03 UTC with strength 1 MPa.}
    \begin{tabular}{cc|cc|cc}
    \hline
    2023 CX1 & RMS (km) & 2024 BX1 & RMS (km) & 2008 TC3 & RMS (km)\\
    \hline
ME (1 MPa; 00 UTC) & 1.26 (1.24)  & ME (1 MPa; 00 UTC)   & 0.56 (0.71)  & ME (0.5 MPa; 03 UTC) & 1.1 (1.2)   \\
ME (5 MPa; 00 UTC) & 0.29 (0.29)  & ME (5 MPa; 00 UTC)   & 0.89 (0.90)  & ME (1 MPa; 03 UTC)   & 1.7 (2.1)   \\
ME (5 MPa; 03 UTC) & 0.74 (0.77)  & -                    & -            & ME (5 MPa; 03 UTC)   & 17.8 (18.4)  \\
ME (1 MPa; 03 UTC) & 0.84 (0.82)  & -                    & -            & -                    & -     \\
H (5 MPa; 00 UTC)  & 1.4 (1.5)    & -                    & -            & -                    & -     \\
\hline
    \end{tabular}
    \label{tab:rms1}
\end{table}

\section{Conclusions}
\label{sec:end}
The fall of small asteroids into the atmosphere a few hours after discovery is an increasingly frequent event, as reported in Table~\ref{tab:small_impact}. The heliocentric orbits of these objects are known from telescopic observations, so the recovery of meteorites on the ground is important to have a complete picture of both the dynamic origin and the chemical-physical composition. Furthermore, immediate recovery of meteorites is important because this minimises contamination from the terrestrial environment, and it is possible to detect the $\gamma-$rays emitted by short-life radiogenic elements generated by the exposure of the meteoroid to cosmic rays, making the extraterrestrial origin certain. However, it is not always possible to triangulate the fireball generated by the asteroid's fall into the atmosphere to locate the strewn field. \\
By analysing the falls of the near-Earth asteroids 2024~BX1, 2023~CX1 and 2008~TC3 with a simple model, which considers a single fragmentation followed by a release of fragments of different mass, all with the same speed and direction with no lateral speed component but with the real wind profile, it was possible to identify the strewn field with uncertainties of the order of a kilometre, a distance easily covered by walking on the ground. The initial data on the asteroid's position and speed at 100 km height derive directly from the heliocentric orbit elements and are available without the need for triangulation of the fireball phase. \\
The most critical parameter to choose has proven to be the average strength $S$ of the asteroid, which conditions the fragmentation height and the path length followed by the fragments and, therefore, the position of the strewn field. The selected values of 0.5, 1 and 5 MPa appear sufficient to have at least one value to give rise to a theoretical strewn field very close to the one actually observed. The equally critical parameter is the mass range of the fragments that ``sample'' the strewn field. Even if it is most reasonable to search for meteorites in the 1 g - 1 kg range, as in our case, because they are most likely to be present, there is no guarantee of that. If there is no fireball data available, the positions of larger meteorites must also be computed because we cannot exclude their existence. Much attention must also be paid to choosing the right atmospheric profile containing the air density, wind speed and direction that can greatly change the position of the strewn field, particularly the smallest fragment. We have seen this high sensitivity to the atmospheric profile in the case of 2024~BX1, which has the meteorite ranges reversed compared to the more standard case of 2023~CX1 or 2008~TC3. The results appear good when the atmospheric profile is computed for the position around the dark flight phase begins. Using an atmospheric profile far from the fall place is acceptable only for a first raw estimate of the dark flight start point. Still, after this preliminary phase, the right profile is mandatory because it can significantly lower the meteorites' search distance from a few kilometres to a few hundred meters. Fortunately, having the right weather profile to compute the position of the possible strewn field is not a problem thanks to the data from soundings, satellites and aircraft that can be integrated to have not only the weather profiles of past events but can also provide those for asteroids a few hours before the impact time. In this way, it is possible to predict where the strewn field will likely be before the asteroid hits Earth. In this sense, it is possible to speak of a strewn field ``ab initio''.\\
Finally, we confirmed that for the fall of the small asteroids 2008~TC3, 2023~CX1 and 2024~BX1, the possible pancake phase after fragmentation is negligible compared to a model with fragmentation alone. Indeed, the results for the position of the strewn field are slightly better without pancakes. Also, the fragmentation number during the fall appears not important, provided that the main body has a final fragmentation.\\

\section*{Acknowledgements}
The authors would like to thank two anonymous referees who, with their suggestions, improved the paper.

\section*{Data Availability}
The data underlying this paper will be shared on reasonable request to the corresponding author. All the software and input data used to describe the 2024~BX1 and 2023~CX1 fall is freely available on GitHub \footnote{\url{https://github.com/AlbinoCarbo/Small-asteroid-fragmentation-model}}. The same software, changing the setting parameters and the atmospheric profile, can be used to describe the 2022~WJ1 and 2008~TC3 falls.
The ESA NEO Coordination Centre orbit determination and impact monitoring Aegis software used for orbit determination is proprietary (\url{https://neo.ssa.esa.int/about-neocc}). 

%-------------------------------------------------------------------
%\bibliographystyle{cas-model2-names}
%\bibliography{Ab_initio_strewn_field}{}
\nocite{*} % to test all bib entrys

\appendix

\section{The strewn field of 2022~WJ1: where are the meteorites?}
\label{sec:appendix}
\begin{figure}
    \centering
    \fbox{\includegraphics[width=\textwidth]{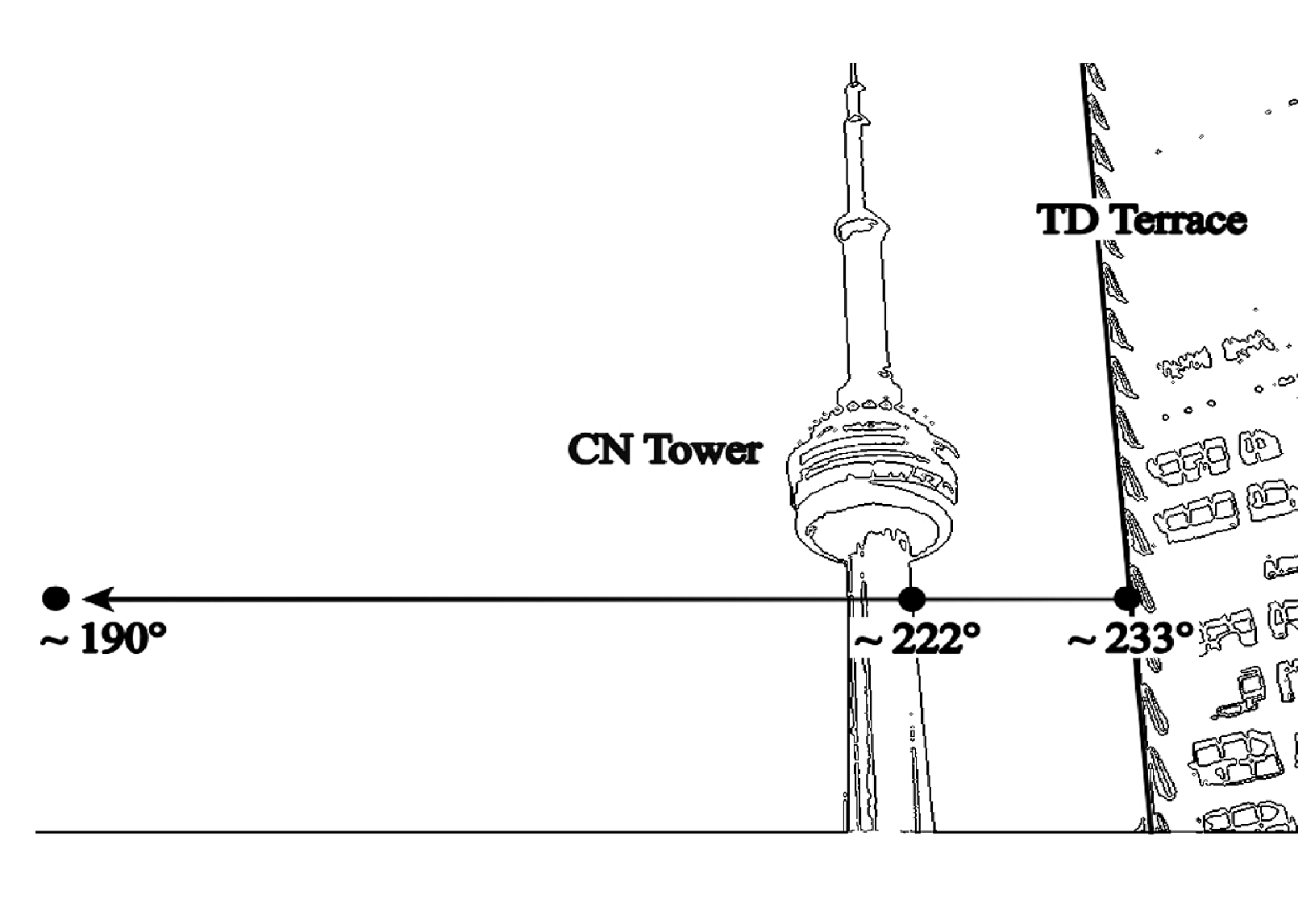}}
    \caption{This qualitative drawing shows the CN Tower and the TD Terrace as seen from the EarthCam, the fireball's trajectory from right to left, and the azimuths of the trajectory determined by the relative dimensions of the tower and the skyscraper as described in the text. The exit point of the fireball from the FOV is that of the video showing the fireball.}
    \label{fig:CN_tower_fireball_trajectory}
\end{figure}

After reproducing with good approximation the strewn fields position of 2023~CX1, 2024~BX1 and 2008~TC3 fall, we tried to apply the same procedure to the case of 2022~WJ1 to determine where the possible meteorites could be. The NEA 2022~WJ1 was discovered by David Rankin with the G96 telescope of the Mt Lemmon Survey on 19 November 2022, at 04:53 UTC, when it appeared as a faint dot of magnitude +19.1. At discovery, the angular speed was 22.6 arcsec/minute, a typical value for a near-Earth asteroid. Rankin followed the object until 05:36 UTC, when the magnitude and the angular speed increased to +18.6 and 35.2 arcsec/minute, respectively. When brightness and angular speed increase significantly in just half an hour, it is reasonable to expect that it is an asteroid already very close to Earth. After the discovery, the near-Earth candidate was included in the NEOCP with the acronym C8FF042. At 06:20 UTC, an email from Richard Kowalski, Lead Survey Operations Specialist of the Catalina Sky Survey, which controls G96, warned the minor planet mailing list that C8FF042 would fall to Earth within a few hours in the Great Lakes Region and that observatories should do everything possible to obtain further astrometric observations before the fall. Seven observatories were able to observe the object before it impacted the Earth's atmosphere at approximately 08:27 UTC over Ontario, Canada.\\
In this region, an EarthCam \footnote{\url{https://www.earthcam.com/world/canada/toronto/cntower/?cam=towerview}} pointed toward the Canadian National Tower (CN Tower) and the Toronto-Dominion Bank Group skyscraper (TD Terrace) in Toronto, captured a video of the fireball created by the fall of C8FF042 into the atmosphere, see Fig.~\ref{fig:CN_tower_fireball_trajectory}. In this video, the white fireball appears to follow a trajectory parallel to the horizon from right to left, and some fragments that detach from the main body are also visible \footnote{\url{https://www.youtube.com/watch?v=i5qxYxSbAzI}}. At 14:19 UTC of the same day, the electronic circular 2022-W69 of the MPC with 46 astrometric observations was released, which assigned the provisional designation 2022~WJ1 to C8FF042, which thus became the sixth asteroid to have hit the Earth a few hours after the discovery.

\begin{figure}
    \centering
    \includegraphics[width=0.8\textwidth]{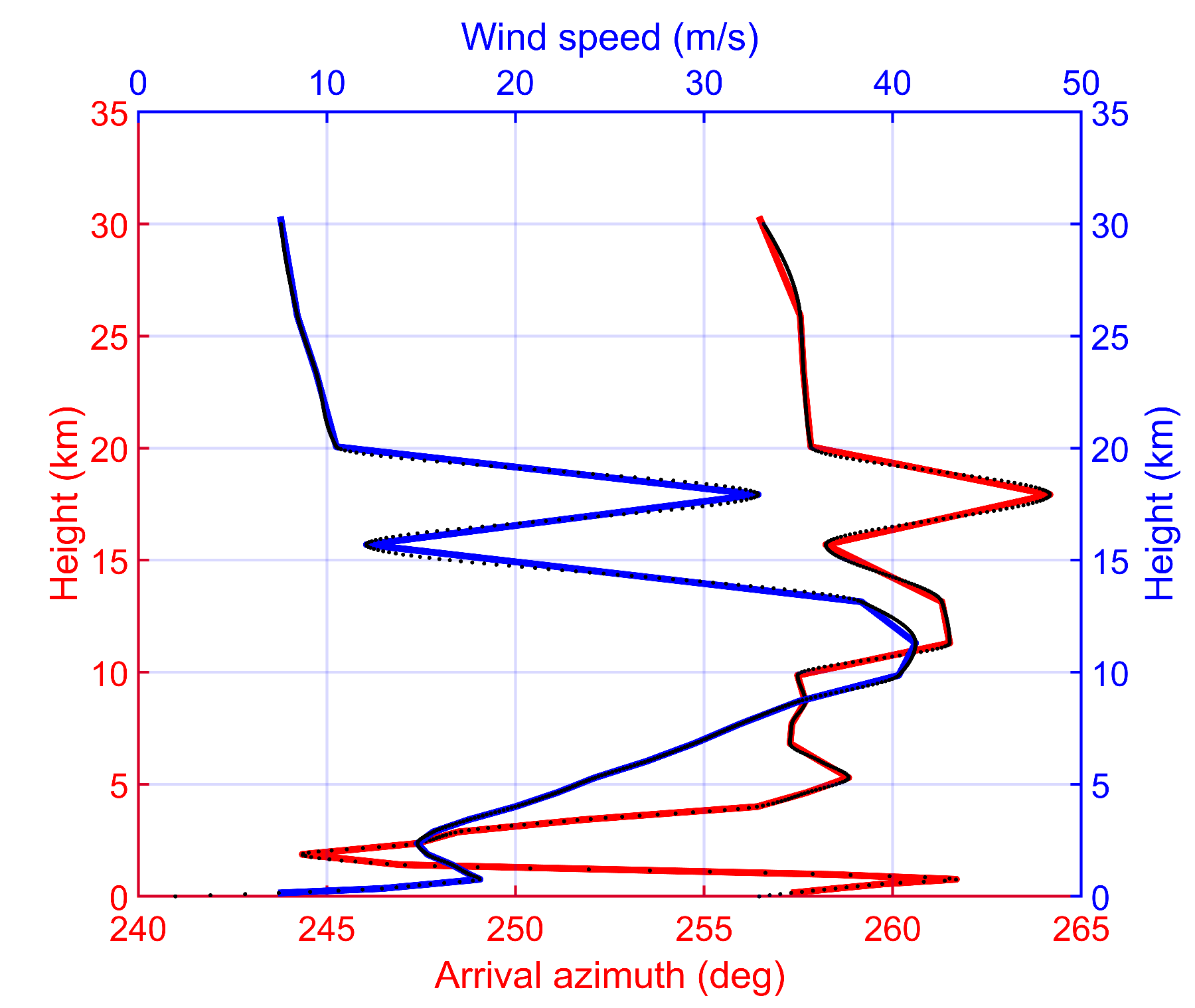}
    \caption{The atmospheric profile used for the 2022~WJ1 event, computed for 19 November 2022, 08 UTC at Lat. $43.2748^\circ$ N and Long. $79.5217^\circ$ W, a point near the start of the dark flight phase for a body of $S\approx 1$ MPa and a fragment of 1 kg final mass. The modulus of the wind speed (m/s) and the arrival azimuth of the wind (deg) are shown. As in previous similar figures, the dotted line interpolates the data provided by the atmospheric profile using a shape-preserving piecewise cubic interpolation.}
    \label{fig:2022WJ1_atmospheric_profile_point1}
\end{figure}

\noindent About 2022~WJ1, we proceeded as in the pilot cases of 2023~CX1, 2024~BX1 and 2008~TC3. The starting conditions upon entry into the atmosphere, computed as in Section~\ref{sec:astrometry_orbits}, are the following (see Table~\ref{tab:orbit2022WJ1}): $h=100$ km, latitude $~\textrm{Lat}_0=43.0887^\circ \pm 0.0003^\circ$ N, longitude $~\textrm{Long}_0=81.7318^\circ \pm 0.0058^\circ$ W, $v_0=13.9576 \pm 0.00012$ km/s, $I_0=22.5267^\circ \pm 0.0038^\circ$ (trajectory inclination) and $A_0=262.6420^\circ \pm 0.0042^\circ$ (fireball's incoming azimuth). As a starting diameter, we had assumed 1 m, with a mean density of $\rho_m\approx 3000 ~\textrm{kg}~\textrm{m}^{-3}$.\\
The radio sounding at 12 UTC from the Buffalo station (ID 72528), 116 km southeast of Brantford, was taken as a first approximation of the atmospheric profile. With this approximate data and strength of 1 MPa, the start of the dark flight for a 1 kg final mass fragment was set at Lat. $43.2748^\circ$ N, Long. $79.5217^\circ$ W (point 1). These coordinates fall in Lake Ontario, and we computed the first atmospheric profile for this point at 08 UTC. Considering the low trajectory inclination above the Earth's surface, which implies a long dark flight trajectory, we have computed a second atmospheric profile for a point at coordinates Lat. $43.2900^\circ$ N and Long. $79.4211^\circ$ W (point 2), intermediate between the approximate start of the dark flight and the approximate falling point of the 1 kg fragment for a strength of 1 MPa. In this way, it was possible to verify that the atmospheric profile does not change significantly with the position along the dark flight path, and the computed strewn fields are virtually the same, with a difference in impact positions of about 200-300 m. For this reason, we only show the strewn field computed with the atmospheric profile for point 1; see Fig.~\ref{fig:2022WJ1_atmospheric_profile_point1}.\\

% Orbital elements table
\begin{table}[!ht]
    \centering
    \caption{Keplerian orbital elements of 2022~WJ1, with the corresponding epoch expressed in MJD, and impact parameters at 100 km altitude from Earth surface. Errors refer to the 1-$\sigma$ formal uncertainties.}
    \begin{tabular}{cccc}
         \hline
      Orbital Data                 & Value                            & Impact Parameter               &  Value \\
         \hline
      Epoch (MJD)                  & 59902.31309 TDT                  & Time (UTC)                     & 2022-11-19 08:26:39.498 $\pm$ 0.052 s       \\ 
      Semi-major axis (au)         &  $   2.1135730  \pm  0.0001926 $ & Latitude (deg)                 & $43.0887  \pm 0.0003$       \\ 
      Eccentricity                 &  $   0.5643474  \pm  0.0000417 $ & East Longitude (deg)           & $-81.7318 \pm 0.0058$        \\ 
      Inclination (deg)            &  $   2.7737726  \pm  0.0001719 $ & Velocity (km s$^{-1}$)         & $13.9576  \pm 0.00012$       \\ 
      Longitude of node (deg)      &  $  56.7047348  \pm  0.0000049 $ & Elevation (deg)                & $-22.5266 \pm 0.00375$       \\ 
      Argument of perihelion (deg) &  $  35.8401362  \pm  0.0002901 $ & Azimuth (deg)                  & $82.6419  \pm 0.00424$       \\ 
      Mean anomaly  (deg)          &  $ 351.3319767  \pm  0.0012654 $ & 1-$\sigma$ semi-major axis (m) & 1314  \\ 
      Normalized RMS               & 0.944                            & 1-$\sigma$ semi-minor axis (m) & 160   \\
         \hline
    \end{tabular}
    \label{tab:orbit2022WJ1}
\end{table}

\begin{figure}
    \centering
    \includegraphics[width=0.8\textwidth]{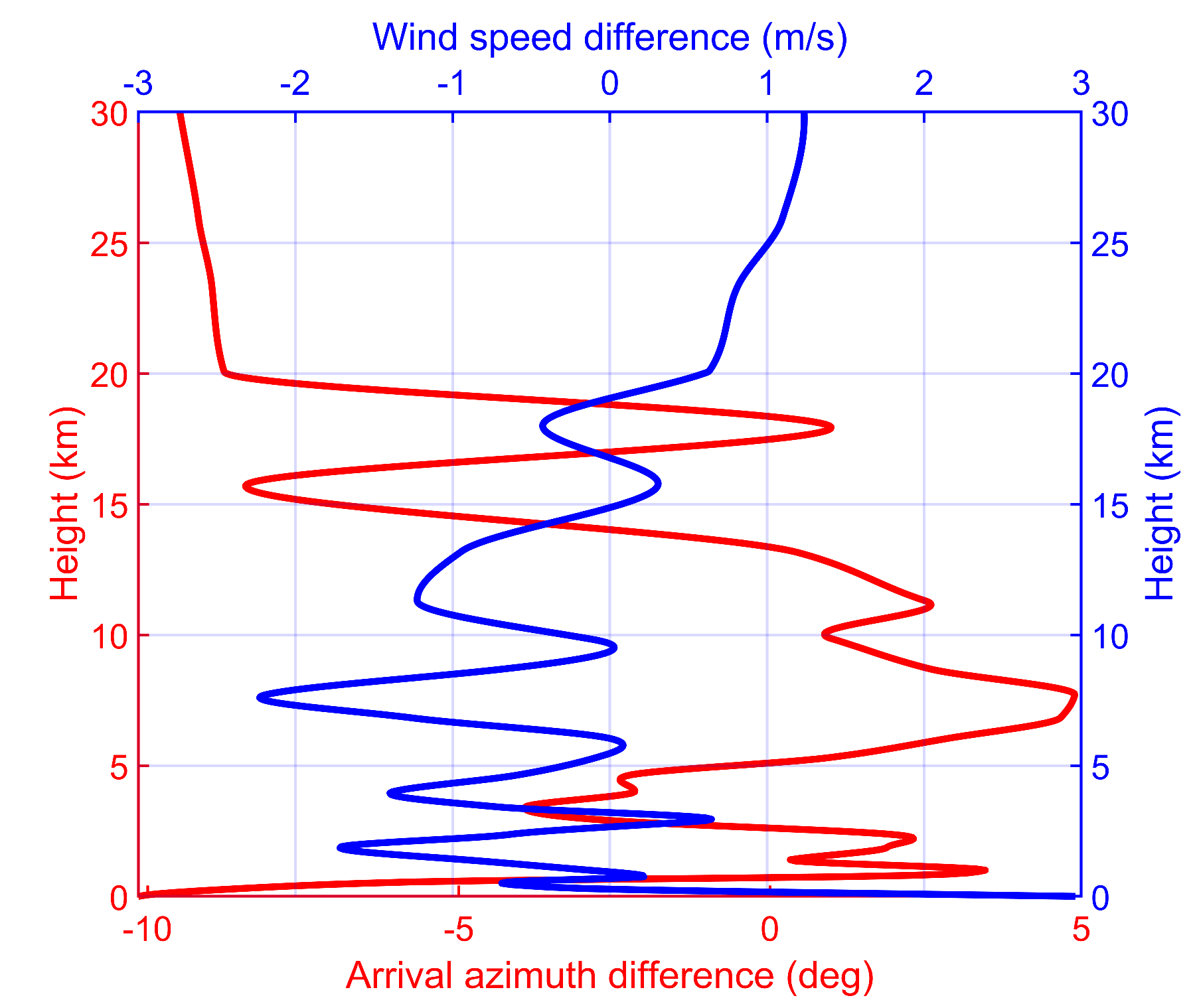}
    \caption{The difference between two atmospheric profiles for the 2022~WJ1 event, computed for Nov 19, 2022, at 08 UTC for Lat. $43.2748^\circ$ N, Long. $79.5217^\circ$ W and Lat. $43.2900^\circ$ N, Long. $79.4211^\circ$ W, same date and hours. The difference in speed is in the order of very few m/s, while the difference in wind azimuth is about $10^\circ$. We have verified that these differences change the strewn field position of 200-300 m, so not in a relevant way.}
    \label{fig:2022WJ1_difference_atmospheric_profile}
\end{figure}

\begin{figure}
    \centering
    \includegraphics[width=\textwidth]{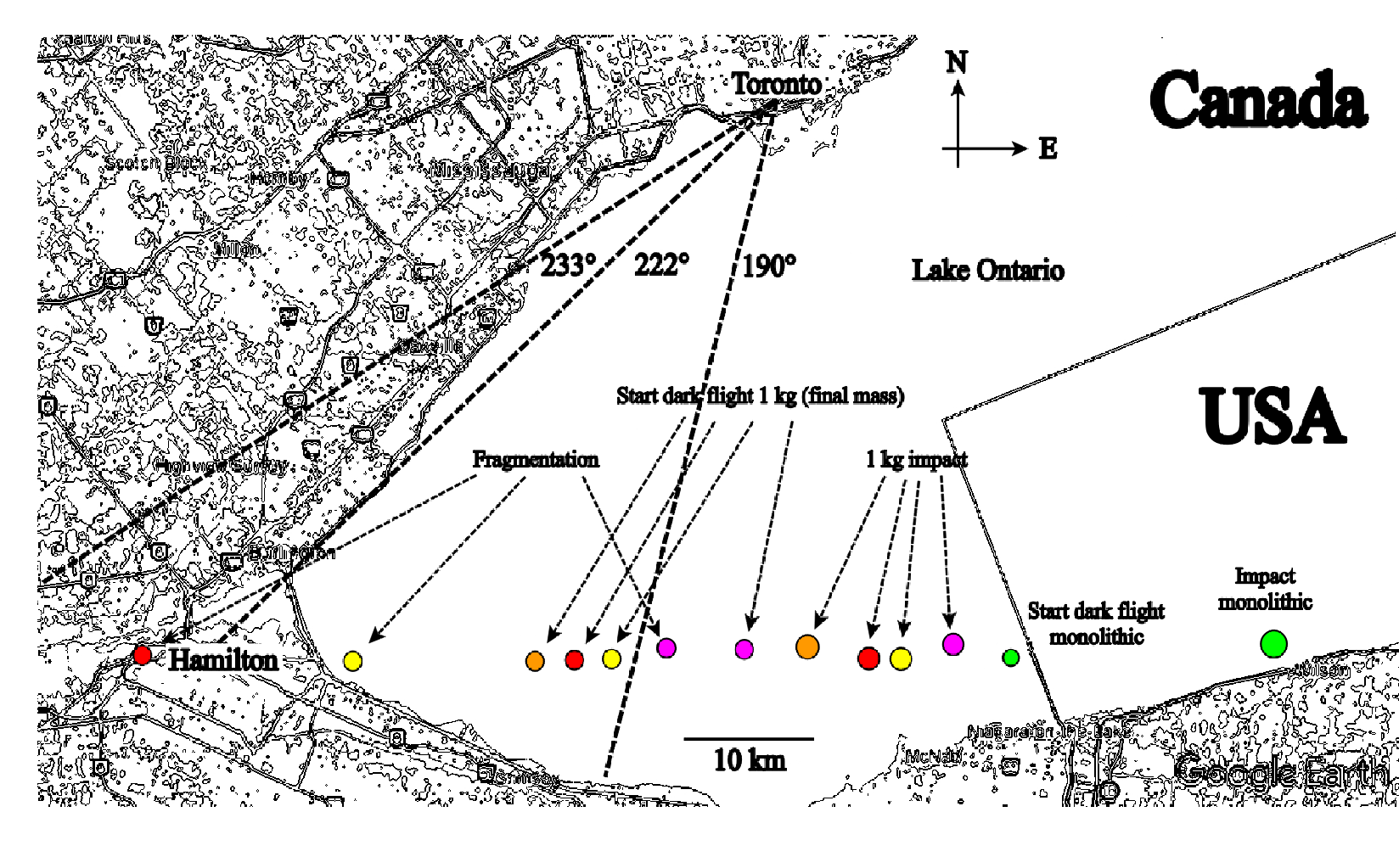}
    \caption{This drawing based on a Google Earth image shows the fragmentation position, start dark flight and impact points for 2022~WJ1. To avoid a too complex drawing, the impact points are those of the fragments with a final mass of 1 kg. Fragments of lesser mass fall first, and fragments of greater mass fall after this point. In any case, they remain aligned in the same east-west direction. The colours indicate different strengths: orange 0.1 MPa, red 0.5 MPa, yellow 1 MPa, pink 3 MPa and green 5 MPa. Also shown are the lines of sight from the EarthCam. The fragmentation for $S=0.1$ MPa is outside the drawing on the left.}
    \label{fig:2022WJ1_strewn_fields}
\end{figure}

\noindent In this case, we computed the possible strewn fields assuming a mean strength of 0.1, 0.5, 1, 3 and 5 MPa. Fragmentation never occurs for strength higher than about 3.6 MPa, so the strewn field for $S=5$ MPa is for a single ablated body. For these possible strewn fields, we computed the position of the 1 kg and the 0.001 kg final mass fragments as extreme limits. As expected, we found that the length increases from 5 to 27 km when the strength decreases from 3 to 0.1 MPa: if the fragmentation occurs at a greater height, the fragments have more time to move away from each other due to drag and atmospheric winds, hence the longer strewn field. The main parameters of these models are reported in Table~\ref{tab:2022WJ1_model} and as can be seen, in all cases, the starting height of the dark flight is less than about 30 km, the height at which the calculated pattern of the atmospheric profile begins. The strewn fields are all aligned (see Fig.~\ref{fig:2022WJ1_strewn_fields}), only shifted some km eastward as the strength increases, and are all 6-10 km away from the lake coast between Grimsby Beach and Port Weller. 
This alignment of the strewn fields is due to the arrival direction of the winds in the troposphere, which had an azimuth between $255^\circ$ and $260^\circ$, see Fig.~\ref{fig:2022WJ1_atmospheric_profile_point1}. This range of azimuth values for the direction of arrival of winds in the troposphere is very near the arrival azimuth of the asteroid. In practice, the falling fragments were pushed from behind, which is the reason for the alignment. Therefore, during the fall, the fragments could not have been blown towards the southern coast of the lake, and this explains why 2022~WJ1 meteorites have never been found: all the fragments that survived the fall into the atmosphere ended up in the lake waters, and none fell on land. \\
The case with $S\approx 5$ MPa differs from the previous one because fragmentation does not occur. In this scenario, assuming a chondrite composition, about 30\% of the original mass survived the ablation phase with a final diameter of about 60-70 cm. Also, this body ended up in the lake but in the United States part, about 1.2 km from Roosevelt Beach: this is the possible impact point closest to the coast. In this case, the fireball phase is also present when the asteroid 2022~WJ1 flies over the lake and ceases in front of Niagara-on-the-Lake, after which the dark flight phase begins. The strewn field estimated by NASA\footnote{\url{https://ares.jsc.nasa.gov/meteorite-falls/events/grimsby-ontario}} for this fall also runs from Grimsby Beach to Port Weller but is much closer to the lake coast and even touches it in some places, but no meteorite was ever found.

\begin{table}
	\centering
	\caption{The main parameters for the different fragmentation models of 2022~WJ1: $S$ is the assumed mean strength of the asteroid, $t_\text{F}$ is the fragmentation time after the start at 100 km height, $h_\text{F}$ is the fragmentation height, $h_\text{DF}$ is the height when the dark flight start for a 1 kg final mass fragment, Lat N/long W DF are the corresponding coordinates, $l_\text{SF}$ is the length of the strewn field between 1 kg and 0.001 kg final mass fragments and Lat N/long W SF are the nominal impact coordinates of the biggest fragment.}
	\label{tab:2022WJ1_model}
	\begin{tabular}{lcccccc} 
		\hline
		$S$ (MPa) & $t_\text{F}$ (s) & $h_\text{F}$ (km) & $h_\text{DF}$ (km) & Lat N/long W DF ($^\circ$) & $l_\text{SF}$ (km) & Lat N/long W SF ($^\circ$)\\
		\hline
        0.1   &  9.4  &  51 & 27 & 43.2741/79.5291 & 27 & 43.2976/79.2537\\
        0.5   &  11.7 &  39 & 26 & 43.2759/79.5052 & 21 & 43.2984/79.2422\\
        1.0   &  12.7 &  34 & 25 & 43.2778/79.4797 & 16 & 43.2993/79.2302\\
        3.0   &  14.8 &  24 & 21 & 43.2861/79.3657 & 5  & 43.3035/79.1695\\
        5.0   &  No   & No  & 13 & 43.3049/79.1005 & 0  & 43.3224/78.8700\\
\hline
	\end{tabular}
\end{table}

\subsection{The fireball of 2022~WJ1 from an EarthCam}
At the end of this Appendix, we want to analyse in more detail the 2022~WJ1 fireball imaged by the EarthCam toward the CN Tower in Toronto. As already mentioned at the beginning of this appendix, on the right side of the FOV (field of view), a very characteristic skyscraper, the TD Terrace, is visible (see Fig.~\ref{fig:CN_tower_fireball_trajectory}). The fireball crosses the FOV with a trajectory parallel to the ground, entering from the right, passing behind the skyscraper and the tower and exiting on the left of the FOV. The fact that the CN Tower and the TD Terrace are simultaneously visible in the image led us to consider the city's northeast quadrant for the EarthCam position. The first problem to solve was identifying the EarthCam's position to determine the approximate trajectory's azimuth range. Unfortunately, requesting the exact location of the cam to the operators did not give the desired results, so we had to proceed with the available data.\\
This problem was solved by measuring the ratio of the angular diameters of TD Terrace and the maximum angular diameter of the bulb of the CN Tower on the image and considering their real dimension with Google Earth. With these data, it is found that the distance $d$ between EarthCam and TD Terrace is given approximately by:

\begin{equation}
d\approx\frac{h \left(l_2/l_1\right)}{\alpha_2/\alpha_1-l_2/l_1}
\label{eq:distance_cam}
\end{equation}

\begin{figure}
    \centering
    \includegraphics[width=\textwidth]{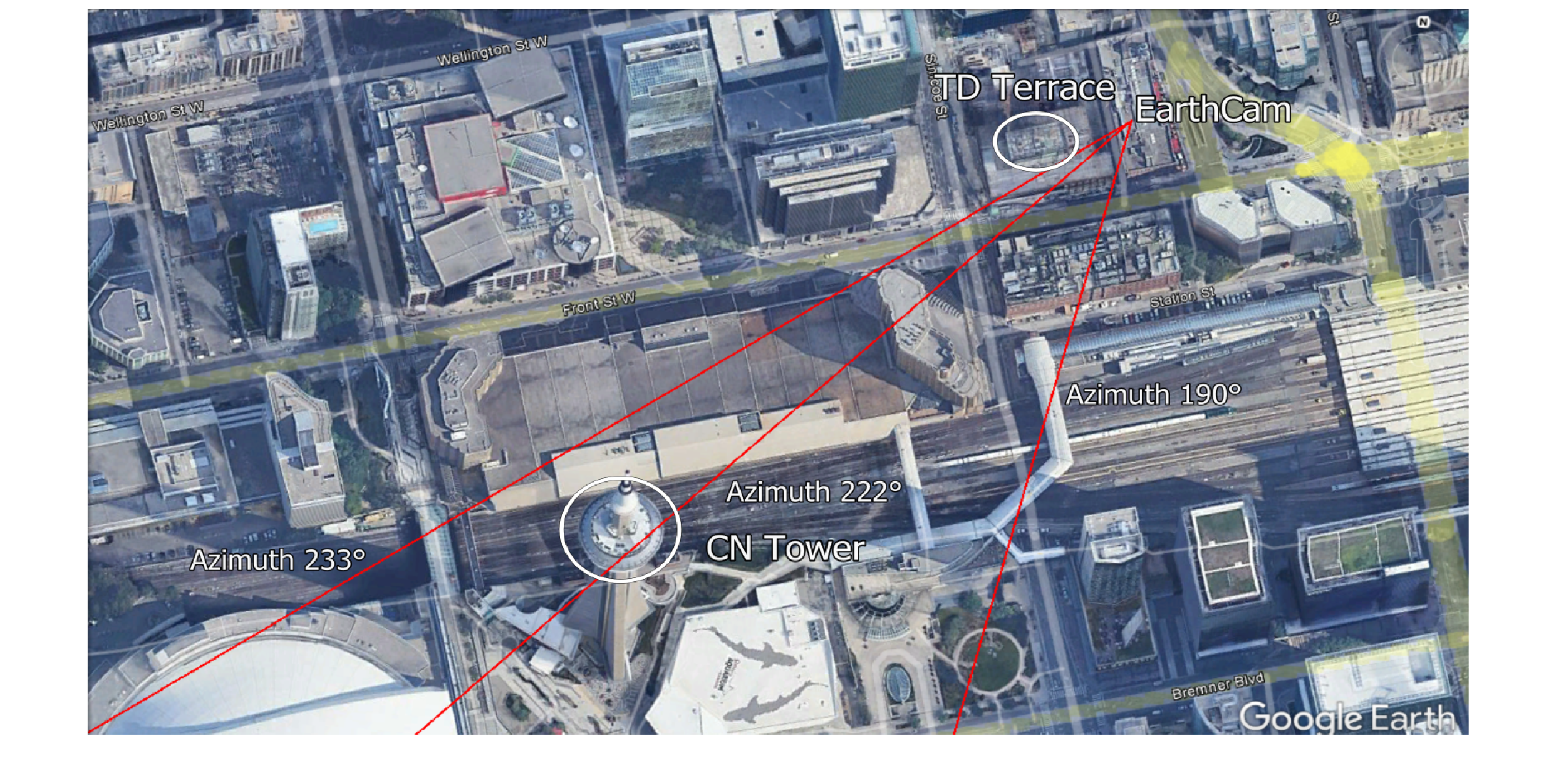}
    \caption{This image, from Google Earth, shows the position of CN Tower, TD Terrace and the EarthCam. The Cam position is the result of our estimate. The lines with the approximate azimuth $233^\circ-222^\circ$ and $190^\circ$ are also drawn. The uncertainty can be estimated in a few degrees. The first two lines delimit the transit of the fireball between the TD Terrace and the CN Tower, while the last line indicates the exit of the fireball from the EarthCam's FOV.}
    \label{fig:CN_tower}
\end{figure}

\noindent In Eq.~(\ref{eq:distance_cam}) $h\approx 360$ m is the distance between TD Terrace and CN Tower along the EarthCam's first approximate line of sight (i.e. north-east), while $l_1\approx 42$ m and $l_2\approx 15$ m are the linear dimension of CN Tower and TD Terrace. Finally, $\alpha_2/\alpha_1\approx 3.9$ is the ratio between the angular dimensions of TD Terrace and CN Tower on the EarthCam image. To estimate the ratio between the angular dimensions, the EarthCam's current daytime video was used, not the one showing the fireball, which was taken at night. The Eq.~(\ref{eq:distance_cam}) is valid if the angular distance between CN Tower and TD Terrace is small and if the facade of the TD Terrace is facing exactly towards the EarthCam. This is not entirely true, but we are interested in an order of magnitude estimate.\\
With these data, from Eq.~(\ref{eq:distance_cam}), we find $d\approx 37$ m from the TD Terrace, which places the EarthCam on the roof of the building opposite the skyscraper, towards the east. So, the distance between EarthCam and CN Tower is about $360 + 37\approx 400$ m. An interesting phenomenon was the reflection of the Sun on the left edge of the skyscraper's glass, visible on 6 May 2024 at 12:33 UTC with a Sun azimuth of about $91^\circ$ and height above the horizon of about $24^\circ$. This phenomenon confirms that the facade of the skyscraper seen by EarthCam faces the east direction. Using the reflection law, it was also possible to estimate the azimuth of the direction in which the EarthCam is located, approximately $60^\circ$ from the north toward the east direction, which defines its approximate position on the roof, see Fig.~\ref{fig:CN_tower}.\\

\noindent Plotting the lines on Google Earth between the estimated position of the EarthCam, the left edge of the TD Terrace and the right edge of the CN Tower pillar (see Fig.~\ref{fig:CN_tower}), we find that the azimuth range between the Tower and the TD Terrace is about $233^\circ-222^\circ$ with an uncertainty that can be estimated at the order of a few degrees. This azimuth range falls on land approximately in correspondence with Hamilton city, in agreement with our fall models of 2022~WJ1 with and without fragmentation, which predicts that 2022~WJ1 is in the fireball phase when in this area. From our model, the height of the fireball in this zone was about 39 km above the ground. Considering the distance of about 60 km from the EarthCam, we can estimate an angular height above the horizon of about $33^\circ$. This is consistent with the elevation above the horizon of the tower bulb as seen from the EarthCam: the bulb is at a mean height of about 350 m from the ground, while our estimated distance CN Tower $-$ EarthCam is about 400 m, considering that the EarthCam is on a roof of about 30 m height, the elevation of the bulb from the EarthCam is about $\arctan\left((350-30)/400\right)\approx 38^\circ$, i.e. the fireball must be some degree lower than bulb as observed, see Fig.~\ref{fig:CN_tower_fireball_trajectory}. The EarthCam records the fireball up to approximately $190^\circ$ azimuth when the body is above the lake in a zone in front of Grimsby Beach, then exits the FOV.\\ 
Considering the models with 0.1 and 5 MPa strengths, the possible strewn field is about 50 km long. Based on previous results on 2024~BX1, 2023~CX1 and 2008~TC3, the cross uncertainties can be estimated in $\pm 1$ km, so there is an area of about $100~\text{km}^2$ to explore, a difficult task even if it was on land. The average depth of Lake Ontario is approximately 86 m, so meteorite recovery is even more difficult.

\end{document}